\documentclass[aps,prx,reprint,amsmath,amssymb,groupaddress,superscriptaddress]{revtex4-1}

\usepackage[utf8]{inputenc}
\usepackage{amsmath,amsfonts,amssymb,mathtools,mathrsfs}
\usepackage{bm}
\usepackage{color}
\usepackage{graphicx}
\usepackage{soul}
\usepackage{CJK}
\usepackage{xcolor}
\usepackage{appendix}

\usepackage{tikz}
\newcommand{\rboxed}[1]{%
    \tikz[baseline=(char.base)]\node[anchor=text, draw, rectangle, rounded corners=2pt, inner sep=2.5pt] (char) {#1};%
}

\newcommand{\bS}{\mathbb{S}}
\newcommand{\Ker}{{\rm Ker\ }}
\newcommand{\dmb}[1]{\textcolor{black}{#1}}

\begin{document}

\begin{CJK*}{UTF8}{gbsn}

\title{Thermodynamic Space of Chemical Reaction Networks}

\author{Shiling Liang (梁师翎)}
\email{shiling.liang@outlook.com}
\affiliation{Institute of Physics, School of Basic Sciences, \'Ecole Polytechnique F\'ed\'erale de Lausanne (EPFL), 1015 Lausanne, Switzerland}
\affiliation{Center for Systems Biology Dresden, 01307 Dresden, Germany}
\affiliation{Max Planck Institute for the Physics of Complex Systems, 01187 Dresden, Germany}
\affiliation{Max Planck Institute of Molecular Cell Biology and Genetics, 01307 Dresden, Germany}
\author{Paolo De Los Rios}
\affiliation{Institute of Physics, School of Basic Sciences, \'Ecole Polytechnique F\'ed\'erale de Lausanne (EPFL), 1015 Lausanne, Switzerland}
\affiliation{Institute of Bioengineering, School of Life Sciences, \'Ecole Polytechnique F\'ed\'erale de Lausanne (EPFL), 1015 Lausanne, Switzerland}
\author{Daniel Maria Busiello}
\email{danielmaria.busiello@unipd.it}
\affiliation{Max Planck Institute for the Physics of Complex Systems, 01187 Dresden, Germany}
\affiliation{Department of Physics and Astronomy ``G. Galilei'', University of Padova, 35131 Padova, Italy}

\date{\today}
\begin{abstract}
Living systems operate out of equilibrium, continuously consuming energy to sustain organised, functional states. Their emergent behaviour usually relies on a set of interconnected chemical reaction networks (CRNs) driven by external fluxes that keep some species at fixed concentrations. Hence, uncovering the principles governing the functioning of these CRNs is crucial to understand how living systems generate and regulate complexity. While kinetics plays a key role in shaping detailed dynamical phenomena, the range of operations of a CRN is fundamentally constrained by thermodynamics. Here, we introduce and analytically derive the ``thermodynamic space" of a CRN, i.e., the range of accessible stationary concentrations that can be realized under a given energetic budget. We establish analogous bounds for reaction affinities, shedding light on how global thermodynamic properties, such as the total non-equilibrium driving, can limit local non-equilibrium quantities, and further showing that these bounds can be used to test candidate network topologies and expose hidden multi-molecular pathways. We illustrate our results in various paradigmatic examples, demonstrating how the onset of complex behaviors is intimately tangled with the presence of non-equilibrium conditions. By providing a general tool for analysing CRNs, the presented framework constitutes a stepping stone to deepen our ability to predict complex out-of-equilibrium phenomena and design artificial chemical systems, starting from the sole knowledge of the underlying thermodynamic properties.
\end{abstract}
\maketitle
\end{CJK*}


\section{Introduction}

Living systems operate far from equilibrium, consuming energy to maintain organized states against the tendency toward increasing entropy \cite{fang2019Nonequilibrium}. This intrinsic non-equilibrium nature of life is sustained by biological processes fundamentally built on chemical reactions \cite{yang2021Physical}. Chemical reaction networks (CRNs) capture many of these biochemical processes, including gene regulation, metabolism, and cellular information processing \cite{zoller2022Eukaryotic,palsson2015Systems,tenwolde2016Fundamental,wong2020Gene}. {Chemical reaction network theory provides} a powerful framework for characterizing CRNs and revealing how complex behaviors emerge from underlying molecular interactions \cite{feinberg2019Foundations,palsson2015Systems}. This complexity manifests in various forms, {such as} non-equilibrium stationary state, multi-stationarity, chaos, and oscillations \cite{vellela2009Stochastic,avanzini2019Thermodynamics,cao2015Freeenergy,gaspard2020Stochastic,li2024Interplaya}. This wide spectrum of possibilities is instrumental for establishing robust biological functions, {while being tangled} with the necessity of continuously harvesting and dissipating energy into the environment.
Indeed, non-equilibrium conditions play a leading role in determining the operations of a biological system, e.g., the abundance of a desired biochemical state can be amplified through energy consumption \cite{li2003Sensitivity,delosrios2014Hsp70,liang2022Emergent,busiello2021dissipation,dass2021Equilibrium,guo2022Nonequilibrium,hathcock2023Nonequilibrium}, free energy dissipation triggers the onset of complex dynamics \cite{zhang2023Free,cao2015Freeenergy}, and chemical driving regulates the emergence of large-scale structures \cite{falasco2018Information,liang2024Thermodynamic,brauns2020PhaseSpace}.

Due to its relevant role, a thermodynamically consistent chemical network theory has been developed to describe the thermodynamics of CRNs --- in and out of equilibrium --- by quantifying the irreversibility of fluxes and relating it to free energy change \cite{pekar2005Thermodynamics,qian2003Stoichiometric,ge2016Nonequilibrium,rao2016Nonequilibrium,maes2021Local}. 
Recent works focused on the role of the topology of CRNs in shaping their non-equilibrium behaviors, studying, for example, how it affects the properties of dissipative fluxes due to external chemostats, i.e., devices that fix the concentrations of selected chemical species \cite{polettini2014Irreversible,dalcengio2023Geometry}, and its role in determining thermodynamic bounds on perturbation responses \cite{owen2020Universal} and characterizing how chemical systems tune their efficiency \cite{bilancioni2025Gearsa}.

Despite the undoubted relevance of understanding out-of-equilibrium phenomena, the general constraints that thermodynamic drivings of any source can impose on biochemical systems are still elusive. In particular, revealing how global non-equilibrium properties constrain specific physical observables and shape the range of operations of CRNs is a formidably hard challenge. 
Recent studies shed some light on the thermodynamic constraints on non-equilibrium amplification, accuracy of information processing, and contrast in reaction-diffusion patterns \cite{maes2013Heat,flatt2023abc,liang2024Thermodynamic,cetiner2022Reformulating,arunachalam2024Informationa}, restricting their analyses to linear (or at most catalytic) CRNs that can always be represented by simple graphs and understood in terms of pseudo-equilibrium quantities \cite{hill1989Free,schnakenberg1976Network}. However, although many biological functions are instantiated by multi-molecular reactions, how these principles and constraints apply to generic CRNs --- described by hypergraphs \cite{klamt2009Hypergraphsa,dalcengio2023Geometry} --- remains an open question.

In this work, we tackle this problem and derive fundamental thermodynamic bounds on reaction affinities and species concentrations valid for any CRN. Our approach goes beyond previous studies by solving the challenging task of considering arbitrary reaction network topologies. This is indeed a crucial step allowing for the application of our results to systems in which complex multi-molecular reactions play a crucial role, such as metabolic pathways, self-assembly processes, and recurrent neural chemical networks \cite{ragazzon2018energy,dack2026Recurrent,ravasio2024minimal}. We introduce the key concept of \textit{thermodynamic space}, that is the accessible region of concentrations within which a biochemical system must operate at stationarity. By providing a general framework for understanding thermodynamic limits of generic CRNs, our work opens new avenues for predicting and analyzing complex behaviors in a wide range of biological and synthetic chemical systems {and provides an inference tool for reaction network topology} \cite{kriebisch2025Roadmap}. As pedagogical yet non-trivial examples, we illustrate the applicability of our findings in a variety of cases, i.e., the bi-stable Schl\"ogl model, a simple self-assembly scheme involving the creation of chemical complexes, a paradigmatic model exhibiting chiral symmetry breaking, and the onset of reaction-diffusion patterns. Finally, to reinforce the broad applicability of our result, we develop TACOS, an open-source Python package to determine the thermodynamic space of any CRN (see App.~\ref{sec:TACOS}).

The paper is structured as follows: Sec.~\ref{sec:outline} presents a detailed outline of main results and take-home messages of this work. In Sec.~\ref{sec:intro_CRN}, we introduce the basic concepts for a thermodynamically-consistent modeling of generic CRNs to characterize their properties and conservation laws. Equipped with these theoretical tools, in Sec.~\ref{sec:main_result}, we present the derivation of our main results, with particular emphasis on the concept of ``chemical probe'' --- an artificial reaction that plays a role analogous to a voltmeter in an electric circuit. This probe allows us to evaluate the thermodynamic accessible space for concentrations of species involved in any possible reaction compatible with the conservation laws. In Sec.~\ref{sec:application}, we apply our results to several representative CRNs to demonstrate their implications and broad applicability. Finally, Sec.~\ref{sec:data} presents two examples in which our bounds can provide useful information from experimental data.

\section{Main results\label{sec:outline}}

A Chemical Reaction Network (CRN) describes a set of interconnected reactions converting chemical species. These reactions can be maintained out of equilibrium through external chemostats, devices that provide continuous fluxes of specific chemical species, keeping their concentrations fixed. Any non-equilibrium CRN is then characterized by its internal species, whose concentrations evolve dynamically, and by external (chemostatted) species, together with the reactions, their associated rates, and a set of conservation laws. Thermodynamics is incorporated into the modeling of CRNs by connecting the ratio of forward and backward fluxes ($J_{\rho\pm}$) of a given reaction $\rho$ to the chemical potentials of the species involved, i.e., to the free energy change along $\rho$, $\Delta_\rho G$. The exact form of this relationship is named local detailed balance (LDB) (Sec.~\ref{sec:dynamiccs_LDB}, Eq.~\eqref{eq:LDB}). Therefore, each reaction $\rho$ is driven by a corresponding thermodynamic force, $A_\rho = RT\ln(J_{\rho+}/J_{\rho-}) = -\Delta_\rho G$, that is termed ``affinity''. This affinity arises from contributions of both internal ($A_\rho^X$) and external ($A_\rho^Y$) species, with $A_\rho^Y$ encoding the non-equilibrium driving (Sec.~\ref{sec:dynamiccs_LDB}, Eq.~\eqref{eq:affinity_decompose}). 

This work presents two principal thermodynamic results that constrain generic CRNs, independently of their structural properties, provided that they reach a stationary state. As such, our framework does not capture the case of persistent oscillations, but it holds for CRNs exhibiting multi-stability. These results connect global network topology and energetic driving to local reaction properties and the system's overall accessible concentration space.

\subsection{Thermodynamic Bounds on Reaction Affinities}

The first main result establishes that the stationary-state affinity of any reaction $\rho$, denoted $A_\rho^{\rm ss}$,
is bounded by global thermodynamic properties of the network. We first decompose the CRN into a minimal set of elementary cycles --- these are the Elementary Flux Modes (EFMs) whose exact definition is introduced in Sec.~\ref{sec:cycle}. Each one of these minimal cycles, denoted by $\bm{e}$, may be driven by chemostatted species and associated with a cycle affinity, $A_{\bm{e}}^Y$, representing the net free energy change induced by these external species (Sec.~\ref{sec:cycle}, Eq.~\eqref{eq:cycle_affinity}). We demonstrate that the affinity $A_\rho^{\rm ss}$ is constrained by the extrema of all affinities associated with elementary cycles that contain $\rho$ (this subset is indicated by $\mathcal{E}_\rho^*$):
\begin{equation}
\left|A_\rho^{\rm ss}\right| \le \max_{\bm{e} \in \mathcal{E}^*_{\rho}} \left|\frac{A_{\bm{e}}^Y}{e_{\rho}} \right| \;,
\label{eq:summary_affinity_bound}
\end{equation}
where each affinity has to be normalized by the effective number of times $\rho$ enters into the corresponding cycle. In Sec.~\ref{sec:main_result}, we derive a sign-resolved form of this bound, which distinguishes whether $\rho$ is accounted in the forward or backward direction, and includes the zero-affinity case (Eq.~\eqref{eq:affinity_bound_1}). In more intuitive terms, the thermodynamic driving force of any local reaction step, $\rho$, is fundamentally limited by a global non-equilibrium driving --- evaluated at the level of the whole CRN --- determining the extremal available affinities along elementary cycles.
These bounds equally hold for effective affinities associated with reactions that are compatible with conservation laws but not explicitly present in the CRN (Sec.~\ref{sec:main_result}, Eq.~\eqref{eq:probe_affinity_bound}).

\begin{figure}[t!]
    \includegraphics[width=1\columnwidth]{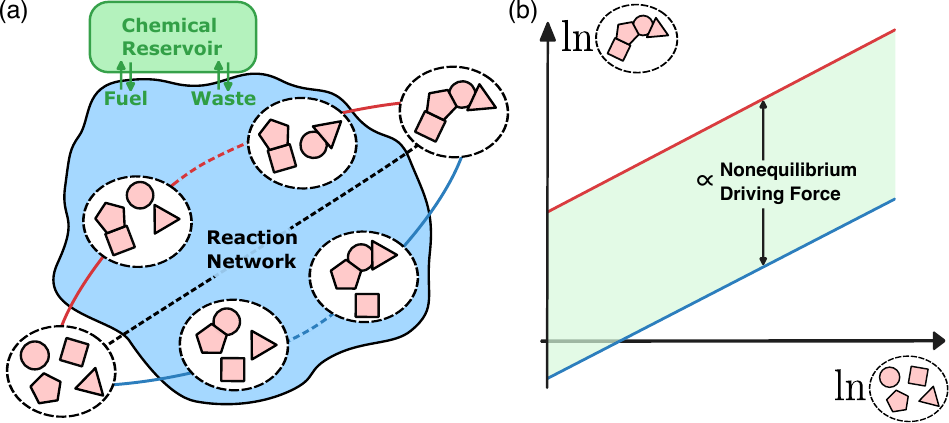}
    \caption{\textbf{Thermodynamic Space of a Chemical Reaction Network (CRN).} (a) Schematic representation of an open CRN where internal species transform and assemble via multiple interconnected pathways, driven by an out-of-equilibrium fuel to waste conversion due to an external chemical reservoir. Highlighted are the extremal pathways {associated with the highest (red) and the lowest (blue) free-energy change}. (b) Conceptual illustration of the thermodynamic space (green region) on logarithmic axes for concentrations. {It shows that the accessible stationary concentrations are bounded from above and below by contributions from the extremal pathways, shown in red and blue in panel (a)}. The width of this space (vertical arrows) quantifies the energy budget via the non-equilibrium driving force.
     }
    \label{fig:thermo_space}
\end{figure}

\subsection{The Thermodynamic Space: Bounds on Accessible Concentrations}

The second main result translates the bound on reaction affinities into constraints on steady-state concentrations of chemical species, hence determining the \textit{thermodynamic space} of the CRN. This is the accessible region of concentrations within which a biochemical system must operate at a given level of non-equilibrium driving.

Consider a net interconversion of internal chemical species --- whether through an existing reaction or a reaction pathway corresponding to an effective reaction compatible with the conservation laws --- such as, for example, the assembly of four different components pictorially represented in Fig.~\ref{fig:thermo_space}a. This interconversion can happen via alternative paths (red and blue pathways in Fig.~\ref{fig:thermo_space}a), and each of them is denoted by $\bm{\pi}^X_{\rho^{+}}$, where $\rho^+$ indicates the specific interconversion involving the set of internal species $X$ from the left to the right side of the reaction (or from the right to the left side for $\bm{\pi}^X_{\rho^{-}}$). The free energy change associated to each path incorporates both the standard Gibbs free energy change associated with the internal species transformation 
and the cumulative Gibbs free energy change induced by the external chemostatted species along that pathway that encode the effect of non-equilibrium drivings. Details on these quantities are provided in Sec.~\ref{sec:main_result}, Eq.~\eqref{eq:affinity_free_energy} and \eqref{eq:eq_constant_bound_0}.
By defining the \emph{effective} equilibrium constant of a reaction pathway as the exponential of (minus) its overall free energy change, and indicating it by $K_{\bm{\pi}^X_{\rho^+}} = {\rm exp}\left(-\Delta_{\bm{\pi}^X_{\rho^+}} G/RT\right)$, we can identify the paths leading to the extremal contributions (these are respectively the red and the blue ones in the sketch of Fig.~\ref{fig:thermo_space}a). We demonstrate that the stationary concentrations $x_i^{\rm ss}$ of the involved species $X_i$ must satisfy the following bounds:
\begin{equation}
K_{\bm{\pi}^X_{\rho^+}}^{\rm min} \le \prod_i (x_i^{\rm ss})^{\bS_{i\rho}^X} \le K_{\bm{\pi}^X_{\rho^+}}^{\rm max} \;,
\label{eq:summary_concentration_bound}
\end{equation}
where $\bS_{i\rho}^X$ are the net stoichiometric coefficients defining the change in internal species $X_i$ for this particular interconversion (Sec.~\ref{sec:intro_CRN}). These constraints dictate that the accessible range of concentrations for a given interconversion is controlled by the global non-equilibrium driving determining the extremal effective equilibrium constants (see the green region in Fig.~\ref{fig:thermo_space}b, delimited by extremal red and blue contributions from respective pathways).

These bounds can be estimated for any set of internal species whose interconversion is compatible with conservation laws. To evaluate them for species not directly linked by an existing reaction, we introduce the idea of the ``chemical probe'' --- an auxiliary, infinitesimally slow reaction introduced to measure the effective affinity between any two sets of species (Sec.~\ref{sec:main_result}\,A). Notice that each of these bounds (upper and lower) can be saturated in generic nonequilibrium conditions when the dynamics is kinetically dominated by the corresponding reaction pathway (red and blue curves in Fig.~\ref{fig:thermo_space}b), i.e., when this pathway is much faster than all the others connecting the same chemical species. {Therefore, Eq.~\eqref{eq:summary_concentration_bound} contains information on the pathways that are more relevant to optimize specific concentrations.}

Writing down Eq.~\eqref{eq:summary_concentration_bound} for all possible interconversions allows one to delineate the entire thermodynamic space for a CRN given a fixed global energy budget.

At thermodynamic equilibrium, all driving forces from external species vanish (e.g., $A_{\bm{e}}^Y=0$ in Eq.~\eqref{eq:summary_affinity_bound}). Consequently, the energy contributions from external species along all pathways are null, causing $K_{\bm{\pi}^X_{\rho^\pm}}^{\rm min}$ and $K_{\bm{\pi}^X_{\rho^\pm}}^{\rm max}$ to coincide and equate to a value solely determined by the standard Gibbs free energy changes of the internal species. Eq.~\eqref{eq:summary_concentration_bound} then becomes an equality, and the system settles into a unique set of concentrations dictated by classical thermodynamics and conservation laws. Conversely, a non-zero width of the thermodynamic space is a signature of non-equilibrium conditions and a prerequisite for the emergence of complex behaviors.

\section{Thermodynamics and Geometry of Chemical Reaction Networks \label{sec:intro_CRN}}

\subsection{Dynamics and Local Detailed Balance (LDB)\label{sec:dynamiccs_LDB}}
\begin{figure*}[!bth]
    \includegraphics[width=1\textwidth]{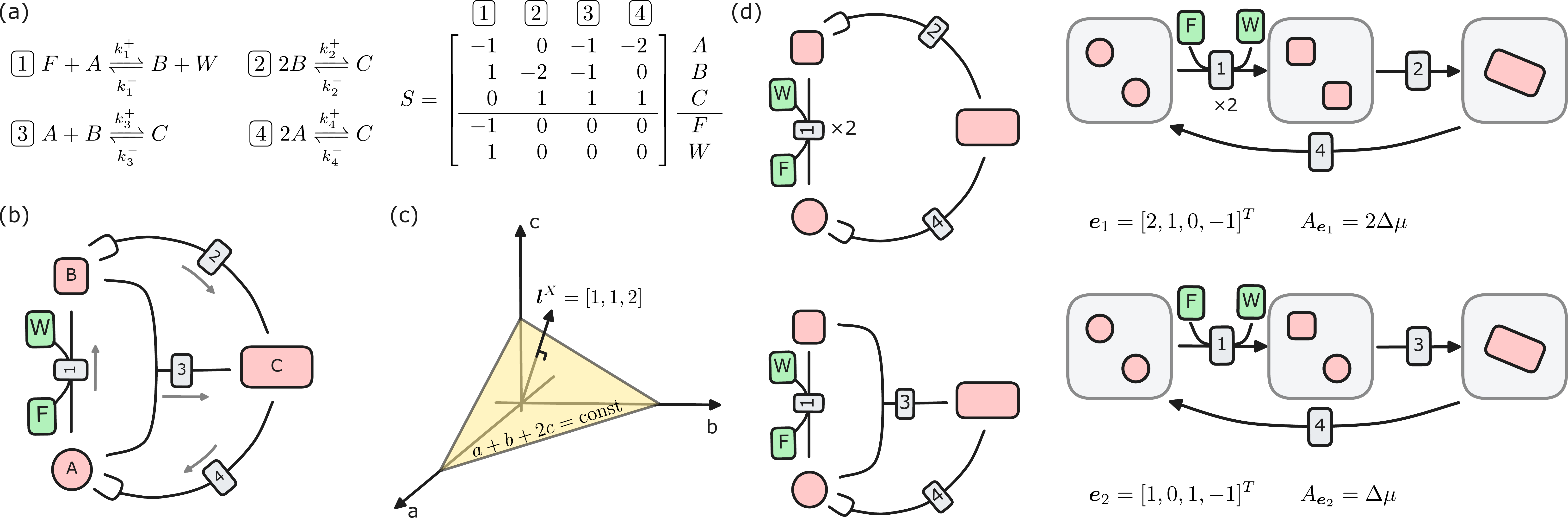}
    \caption{
    \textbf{An example Chemical Reaction Network (CRN).} (a) {An illustrative} set of chemical reactions with their associated rates and the corresponding stoichiometric matrix. These quantities define the CRN. $A$ and $B$ denote two different forms of a monomer (e.g., inactive and active), while $C$ is a dimer. $F$ and $W$ are the chemostatted external species. A chemical potential difference $\Delta\mu = \mu_F - \mu_W$ drives the conversion of $A$ into $B$, pushing the system out of equilibrium. (b) The hypergraph representation of the reaction network. Here, the number of edges pointing to each chemical species represents its stoichiometric coefficient in the reaction. {The arrows indicate the orientation of each reaction edge, as determined by the signs of the corresponding entries in the stoichiometric matrix, pointing from {reactants} with negative entries to products with positive entries.} (c) The conservation law $\bm{l}^X$ is derived from the stoichiometric matrix $\bS^X$ and is orthogonal to the stoichiometric subspace of internal species. As in the main text, concentrations are indicated with lowercase letters. (d) Two driven elementary flux modes (EFMs) involving reaction 4. An EFM represents a cyclic reaction pathway that does not result in a net change of internal species. EFMs are represented by vectors $\bm{e}$, whose entries indicate the number of times each reaction occurs within the cycle, with the sign indicated the direction in which the reaction is performed (Sec.~\ref{sec:cycle}). The cycle affinity of an EFM, $A_{\bm{e}}$, corresponds to the net chemical potential difference associated with external species consumed or produced along the cycle.}
    \label{fig:CRN_intro}
\end{figure*}

Here, we first summarize the relevant aspects of the framework to characterize dynamics and thermodynamics of open CRNs that are maintained out of equilibrium through the presence of chemostats.
We categorize the chemical species $\{Z_\sigma\}$ into two distinct types: internal species $\{X_i\}_{i = 1, \dots N_X}$, whose concentrations are governed by the reaction dynamics within the network, and external (chemostatted) species $\{Y_j\}_{j = 1, \dots N_Y}$, which are maintained at constant concentrations by the action of chemostats. Let $\rho$ denote the index of one of the $N_R$ possible reactions in the CRN. We can represent a generic reaction in the following form:
\begin{equation}\label{eq:reaction}
\sum_{j=1}^{N_Y} \nu_{j,\rho}^{Y+} Y_j+\sum_{i=1}^{N_X}\nu_{i,\rho}^{X+}  X_i  \xrightleftharpoons[k_{\rho}^-]{k_{\rho}^+}\sum_{j=1}^{N_Y} \nu_{j,\rho}^{Y-} Y_j+\sum_{i=1}^{N_X}\nu_{i,\rho}^{X-}  X_i  ,
\end{equation}
where $\nu_{i,\rho}^{X+}$, $\nu_{i,\rho}^{X-}$ denote the stoichiometric coefficients of the internal species $X_i$ in the forward and reverse reactions, respectively. Here, $\nu_{j,\rho}^{Y+}$, $\nu_{j,\rho}^{Y-}$ represent the same quantities for the external species $Y_j$. The rate constants for the forward and reverse reaction are respectively given by $k_{\rho}^+$ and $k_{\rho}^-$. As discussed later, to ensure thermodynamic consistency, we consider all chemical reactions to be bidirectional, i.e., reversible CRNs. The stoichiometric changes induced by all reactions in the network are captured by the stoichiometric matrix $\bS$:
\begin{equation}\
   \bS= 
   \left(
  \begin{array}{cc}
    \text{internal species} \\\hline
    \text{external species} \\
    \end{array}
\right)
  =\left(
  \begin{array}{cc}
    \bS^X \\\hline
    \bS^Y \\
  \end{array}
\right)
\end{equation}
where $\bS^X$ and $\bS^Y$ are two block matrices whose entries are defined by $\bS_{i,\rho}^X = \nu_{i,\rho}^{X-} - \nu_{i,\rho}^{X+}$ and $\bS_{j,\rho}^Y = \nu_{j,\rho}^{Y-} - \nu_{j,\rho}^{Y+}$, representing the net change in molecular numbers for internal and external species, respectively, due to the $\rho$-th reaction. $\bS^X$ has dimensions $N_X \times N_R$, while $\bS^Y$ is a $N_Y \times N_R$ matrix. For the sake of generality, $Z_\sigma$ denotes a generic chemical species (internal or external) and $\bS_{\sigma,\rho}$ the entry of $\bS$ corresponding to the change of its molecular number due to the $\rho$-th reaction. The $\rho$-th column of $\bS$, denoted as $\bS_\rho$, represents the change in molecular numbers for all species in the $\rho$-th reaction. In Fig.~\ref{fig:CRN_intro}a, we present the example of a simple CRN characterized by four reactions among 5 chemical species, 3 internal, $\{X\}=\{A, B, C\}$, and 2 external, $\{Y\}=\{F, W\}$. Its stoichiometric matrix is also reported in the figure.

For dilute CRNs, the concentration evolution can be expressed through the mass-action law. In particular, we can introduce the forward and backward flux vectors, ${\bm J}^+$ and ${\bm J}^-$, whose $\rho$-th component quantifies the propensity of a given reaction $\rho$ in the forward or backward direction, respectively. The explicit form of these fluxes follows the mass-action kinetics:
\begin{equation}
    J^+_\rho = k^+_\rho \prod_{i=1}^{N_X} {[X_i]}^{\nu^{X+}_{i,\rho}} \prod_{j=1}^{N_Y} {[Y_j]}^{\nu^{Y+}_{j,\rho}}
    \label{massaction}
\end{equation}
and analogously for $J^-_\rho$, where $[X_\sigma]$ indicates the concentration of the species $X_\sigma$. By further introducing the net flux vector $\bm{J}$, whose elements are $J_\rho = J_\rho^+ - J_\rho^-$, we can express the dynamics of species concentrations in the network through the stoichiometric matrix as:
\begin{equation}\label{eq:reaction_dynamics}
    \frac{d\bm{x}}{dt} = \bS^X\bm{J},\quad \frac{d\bm{y}}{dt} = \bS^Y\bm{J}+\bm{I}^Y=\bm{0} \;.
\end{equation}
Here, we used the short-hand notation $\bm{x}$ to indicate the vector of concentrations of internal species, so that its element $x_\sigma = [X_\sigma]$, and analogously for $\bm{y}$. We denote the complete set of species concentrations as $\bm{z} = (\bm{x},\bm{y})$. The external flux $\bm{I}^Y$ is induced by chemostats and maintains constant concentrations of chemostatted species by compensating the flux due to internal reactions. If the system reaches a steady state, it will satisfy the condition
\begin{equation}\label{eq:steady_state}
\bS^X \bm{J}^{\rm ss}=\bm{0},
\end{equation}
where $\bm{J}^{\rm ss}$ represents the steady-state flux vector, i.e., the flux where all concentrations have been replaced by their stationary values. However, it is important to note that not all CRNs necessarily reach a steady state. Some systems may exhibit limit cycles, chaos, or other types of attractors. In this work, we restrict our discussion to the CRNs that can reach stationary state solutions.

Thermodynamics is incorporated through the local detailed balance (LDB) condition \cite{rao2016Nonequilibrium}, which constrains the ratio of forward to backward fluxes:
\begin{equation}\label{eq:LDB}
\frac{J_\rho^+}{J_\rho^-} = e^{-\Delta_\rho G /RT} = e^{A_\rho /RT},
\end{equation}
where $\Delta_\rho G = \sum_{\sigma}\mu_{Z_\sigma}\bS_{\sigma,\rho}$ is the free energy change along the reaction $\rho$, and $A_\rho = -\Delta_\rho G$ is the reaction affinity. The affinity aligns with the direction of the net reactive flux $J_\rho$, i.e., a positive $A_\rho$ is associated with a positive $J_\rho$ since it indicates that the forward flux is greater than the backward one.
The chemical potential of a species $Z_\sigma$ is given by $\mu_{Z_\sigma} = \mu_{Z_\sigma}^{\circ} + RT \ln z_\sigma$, where $\mu_{Z_\sigma}^{\circ}$ is the standard chemical potential and $RT \ln z_\sigma$ is the entropic contribution, again recalling that $z_\sigma = [Z_\sigma]$. We then introduce the chemical potential vector $\bm{\mu}$ whose elements are $\mu_\sigma = \mu_{Z_\sigma}$.
By employing this simplified notation, the affinity vector for all reactions in the network can be computed as:
\begin{equation}\label{eq:affinity_vector}
\bm{A} = -\bS^T \bm{\mu} \;,
\end{equation}
where $\bS^T$ is the transpose of the stoichiometric matrix.
By splitting the chemical potential vector into the contributions from internal and external species, $\bm{\mu}  = (\bm{\mu}^X,\bm{\mu}^Y)$, the affinity can be decomposed as:
\begin{equation}\label{eq:affinity_decompose}
    \bm{A} = \bm{A}^X+\bm{A}^Y = - (\bS^X)^T\bm{\mu}^X-(\bS^Y)^T\bm{\mu}^Y.
\end{equation}
The affinity $\bm{A}^X$ represents the thermodynamic driving force associated with changes in internal species, while $\bm{A}^Y$ represents the thermodynamic driving force associated with changes in external species.

By taking advantage of the mass-action kinetics in Eq.~\eqref{massaction}, the LDB condition can be rewritten in terms of reaction rate constants:
\begin{equation}
\frac{k_\rho^+}{k_\rho^-} = e^{-\Delta_\rho G^{\circ}/RT},
\end{equation}
where $\Delta_\rho G^{\circ} = \sum_\sigma \mu_{Z_\sigma}^{\circ} \bS_{\sigma,\rho}$ is the change in the standard free energy along the reaction $\rho$. This form corresponds to the LDB commonly referred to in stochastic thermodynamics \cite{maes2021Local}, even if the expression in Eq.~\eqref{eq:LDB} can be considered to be valid more in general. We also remark that our results are solely based on Eq.~\eqref{eq:LDB} and not on the specific form of the kinetics.

\subsection{\label{sec:cycle}Cycles and Elementary Flux Modes (EFMs)}
A generic CRN can be regarded as a hypergraph in which each reaction is a hyperedge connecting multiple reaction species \cite{bretto2013hypergraph,dalcengio2023Geometry}. The stoichiometric matrix defines the geometry of the hypergraph, from which we can extract the hypercycles as a sequence of reactions through which the system remains unchanged. From an algebraic perspective, a vector $\bm{v}$ is a cycle of a CRN, i.e., it contains the set of reactions that produces no net changes in molecular numbers, if $\bS \bm{v}=\bm{0}$, namely the vector belongs to the right nullspace of the stoichiometric matrix, also called the kernel of the matrix, $\Ker(\bS)$. In Fig.~\ref{fig:CRN_intro}b, we show the hypergraph associated with the stoichiometric matrix of the example in Fig.~\ref{fig:CRN_intro}a.

In an open CRN, external species are chemostatted while internal species are affected by the reaction dynamics. We are thus interested in the cycles of the stoichiometric matrix of internal reactions, $\bS^X$. These cycles $\bm{c}$, named ``flux modes'', are vectors of size $N_R$ and satisfy 
\begin{equation}
    \bS^X\bm{c}=\bm{0} \;.
\end{equation}
This equation is solved by all possible sets of reactions that preserve the molecular numbers of internal species only. The space constituted by all flux modes is the kernel of the internal stoichiometric matrix, $\Ker(\bS^X)$. We can immediately see that, according to Eq.~\eqref{eq:steady_state}, the stationary solution $\bm{J}^{\rm ss}$ of a CRN is a flux mode.
That means we can interpret the flux modes as all possible steady-state solutions for the fluxes, while the actual solution depends on the specific values of the rates.

Understanding all possible pathways through which a CRN can operate is of fundamental importance to introduce the concept of ``elementary flux modes'' (EFMs). 
{EFMs are support-minimal non-zero flux modes in $\Ker(\bS^X)$, and the full set of EFMs forms a generating set for this kernel, in the sense} that all possible flux modes in the CRN can be expressed as linear combinations of EFMs. {More precisely,} an EFM is a vector $\bm e$ of size $N_R$ whose support, i.e., the set of indices of non-zero entries ${\rm supp}(\bm{e}) = \{\rho\ |\ e_\rho\neq0\}$, is minimal among non-zero vectors in $\Ker(\bS^X)$. As such, an EFM cannot be further decomposed into flux modes with strictly smaller support. In other words, EFMs are minimal cycles preserving the molecular numbers of internal species. The set of all EFMs is indicated by $\mathscr{E}$.
We indicate with $e_\rho$ the $\rho$-th element of the EFM $\bm{e}$. The modulus of $e_\rho$ represents the number of times the reaction $\rho$ occurs along $\bm{e}$. However, the sign of $e_\rho$ indicates whether the reaction is performed in the forward (positive) or backward (negative) direction in the EFM. Notice that this definition intrinsically depends on the arbitrary choice of the positive direction of a given reaction $\rho$ that we consider to be the forward one and associate with a rate $k^+_\rho$. A different choice will lead to the same result.

The relationship between EFMs and flux modes can be characterized through the concept of conformality. We say that an EFM $\bm{e}$ is conformal to a flux mode $\bm{c}$ if, for each reaction $\rho$, either $e_\rho = 0$ or $e_\rho c_\rho > 0$. This property ensures alignment between the EFM and the flux mode whenever they share a reaction. A fundamental property of EFMs is that any flux mode admits a conformal decomposition, i.e., it can be written as a sum without {cancellation of EFMs \cite{muller2016Elementary,bilancioni2025Gearsa}, each of them conformal to it. Such a decomposition is generally not unique, and different flux modes may involve different subsets of the full set $\mathscr{E}$.} Consequently, since the steady-state flux $\bm{J}^{\rm ss}$ is a flux mode, it also admits a conformal decomposition. In Section \ref{sec:main_result}, we will leverage this property to prove our main results.

Here we also briefly comment on the numerical enumeration of EFMs. For fully irreversible networks, EFMs coincide with the extreme rays of the non-negative steady-state flux cone defined by the internal stoichiometric matrix \cite{schuster1994ELEMENTARY}. In the reversible convention adopted in this work, we instead introduce a split representation $\bS^{X,\text{split}}\equiv[\bS^X,-\bS^X]$ and compute extreme rays in split space, which are then projected back to the original reaction coordinates. A detailed description of these algorithms and their scaling properties with the network size is provided in Appendix~\ref{sec:EFM_enumeration}. {It is worth noting that, while for small to medium CRNs, these algorithms can provide a practical tool to identify EFMs, larger networks would require more targeted computational tasks. In what follows, we will comment on how the presented results can guide the development of such targeted search.}

\subsection{\label{sec:conservation_law}Conservation laws and Reaction space}
Another crucial ingredient to fully characterize a CRN is the identification of conservation laws. They are encapsulated in the left null space of $\bS$, known as the cokernel of $\bS$ or equivalently the kernel of $\bS^T$\cite{palsson2015Systems,rao2016Nonequilibrium}. We denote the set of all conservation laws as $\mathscr{L}=\Ker[\bS^T]$, which is also named ``metabolic pool'' in the context of metabolic networks \cite{famili2003Convex}. Hence, for any conservation law $\bm{l}\in\mathscr{L}$ of a CRN, $\bm{l}^T \bS=\bm{0}$ holds. In an open CRN, the introduction of chemostats breaks some conservation laws, particularly those involving chemostatted species $\{Y_j\}_{j=1,\dots N_Y}$ \cite{polettini2014Irreversible}. Consequently, only a subspace of $\mathscr{L}$ represents conserved quantities in an open system. This subspace, denoted  by $\mathscr{L}^X$ is exactly the cokernel of $\bS^X$, i.e., $\mathscr{L}^X = \Ker[(\bS^X)^T]$. For any $\bm{l}^X\in \mathscr{L}^X$, we can identify a corresponding conserved quantity $(\bm{l}^X)^T\bm{x}$:
\begin{equation}
    \frac{d(\bm{l}^X)^T\bm{x}}{dt} =  (\bm{l}^X)^T\bS^X\bm{J} = \bm{0} \;.
\end{equation}
Figure \ref{fig:CRN_intro}c reports the internal conservation law associated with the example reaction network in Fig.~\ref{fig:CRN_intro}a. Although some conservation laws are violated in the presence of chemostatted species, the entire set $\mathscr{L}$ remains a valuable tool for identifying permissible reactions in a given CRN. These are reactions that can happen in principle, since they are compatible with the conservation laws determined by $\bS$, even if they are not possible in the CRN under investigation. Analogously to the concept of EFMs, we introduce the set of elementary conservation laws (ECLs). Each ECL $\bm{r}$ has a minimal and irreducible support. Thus, each feasible reaction not present in a given open CRN, $\hat{\rho}$, is represented by stoichiometric columns $\bS_{\hat{\rho}}$, satisfying $\bm{r}^T\bS_{\hat{\rho}} = 0$ for all $\bm{r}$. We name ``reaction space'', $\mathscr{R}$, the set of all possible reactions in a CRN that are compatible with the conservation laws associated with the full stoichiometric matrix $\bS$.

The reaction space will be the foundational point to determine the thermodynamic space of general CRNs. In fact, the introduction of a fictitious test reaction will allow us to bound the range of concentrations by exploring also chemical transitions that are possible --- but not present --- in a CRN, i.e., those belonging to $\mathscr{R}$.


\subsection{Global thermodynamics of CRNs - Cycle affinity and Equilibrium condition}
The thermodynamic properties of CRNs are fundamentally characterized by the LDB condition, Eq.~\eqref{eq:LDB}. This condition relates the degree of irreversibility of reaction fluxes to changes in Gibbs free energy, which is equal to (minus) the affinity of the reaction, and represents a local relationship. Additionally, CRNs have global thermodynamic properties that are associated with their topology and captured by the EFMs.

For any given EFM, $\bm{e}\in\mathscr{E}$, we can evaluate a cycle affinity as follows:
\begin{equation}\label{eq:cycle_affinity}
A_{\bm{e}} = \bm{e}^T \bm{A} = \underbrace{\bm{e}^T \bm{A}^X}_{=0}+ \underbrace{\bm{e}^T \bm{A}^Y}_{=-\Delta_{\bm{e}} G_Y}  = A^Y_{\bm{e}}.
\end{equation}
which is the net affinity considering all reactions participating to $\bm{e}$. Notice that Eq.~\eqref{eq:cycle_affinity} holds for both stationary and non-stationary states \cite{beard2004Thermodynamic,de1936thermodynamic}. By definition, proceeding along an EFM does not alter internal species concentrations, hence $\bm{e}^T \bm{A}^X = 0$. The cycle affinity arises solely from concentration changes of chemostatted species, as they are compensated by external fluxes (see Eq.~\eqref{eq:reaction_dynamics}). Consequently, the cycle affinity equals the negative Gibbs free energy change in the chemical reservoirs performing the corresponding reaction cycle, i.e., $A^Y_{\bm{e}}=-\Delta_{\bm{e}} G_Y$. This is a sheer structural property of the CRN and is independent of the concentrations of the internal species \cite{schnakenberg1976Network}. Figure \ref{fig:CRN_intro}d indicates the EFMs of the network in Fig.~\ref{fig:CRN_intro}a that involve reaction $4$. We also report the cycle affinity associated with each EFM.

It is crucial to understand how this property reflects the equilibrium and non-equilibrium conditions of a CRN. At equilibrium, all net fluxes vanish, i.e., $\bm{J}^{\rm eq}=\bm{0}$, and accordingly, all affinities are equal to zero, i.e., $\bm{A}^{\rm eq}=\bm{0}$. This means that, by using the definition of affinity vector in Eq.~\eqref{eq:affinity_vector}, the equilibrium condition equivalently requires $\bS^T\bm{\mu}^{\rm eq} = \bm{0}$. Closed CRNs always reach equilibrium. On the contrary, open CRNs often cannot equilibrate, as $\bm{A}^{\rm eq}=\bm{0}$ is generally incompatible with the cycle affinity condition in Eq.~\eqref{eq:cycle_affinity}. Indeed, the presence of chemostatted species typically maintains open CRNs out-of-equilibrium, potentially leading to the emergence of a variety of attractors (e.g., fixed points, limit cycles, strange attractors) depending on the kinetics. However, if all chemostatted species are maintained at their equilibrium concentrations, i.e., $\bm{e}^T \bm{A}^Y=0$ for each EFM $\bm{e}$, the open CRN will relax to equilibrium \cite{schuster1989Generalization, rao2016Nonequilibrium}.

\section{Universal Thermodynamic Bounds for chemical reaction networks}\label{sec:main_result}
\subsection{Thermodynamic accessible region for affinities}
Cycle affinities along EFMs offer a global perspective on the thermodynamic features of CRNs and quantify their deviation from equilibrium conditions. However, the steady-state values of all reaction affinities throughout the network remain challenging to determine. In fact, the affinity distribution stems from the presence of external chemostatted species and its determination requires solving the system with detailed kinetics. The inherent non-linearity of most non-equilibrium CRNs often precludes analytical solutions, making necessary the use of numerical methods to characterize the thermodynamics of non-equilibrium stationary states. Despite the dependence of the affinity distribution on kinetic details, here we demonstrate that thermodynamic bounds on reaction affinities, that are local quantities, can be established using only global thermodynamic and topological properties of the CRN. This approach provides valuable insights into the behavior of non-equilibrium systems without requiring complete knowledge of their kinetics. Moreover, it offers a promising perspective on the understanding of how --- and within which limits --- thermodynamic drivings can effectively enhance (or reduce) reaction affinities, an effect shown to be relevant in various chemical systems \cite{delosrios2014Hsp70,hartich2015Nonequilibrium,chen2020Enhanceda}. The main result of this part is summarized in Sec.~\ref{sec:outline}~A, and below we present its derivation.

\subsection*{Derivation for reactions in the CRN}

To derive these universal thermodynamic bounds, we begin by assuming the existence of (one or multiple) steady-state solutions with non-zero species concentrations for the CRN, represented by fixed points of Eq.~\eqref{eq:reaction_dynamics}.
We recall from Sec.~\ref{sec:cycle} that, since $\bm{J}^{\rm ss}$ admits a conformal decomposition, every non-zero flux along an edge in the CRN --- equivalently, every reaction flux $J_\rho$ --- must be contained in at least one EFM that is oriented in the same direction along that edge, i.e., a conformal EFM. 
Moreover, for a given reaction $\rho$, the component of the flux $J_\rho$ has the same sign as the affinity $A_\rho = RT\ln(J^+_\rho/J^-_\rho)$. Thus, conformal EFMs are also aligned with the steady-state affinity vector $\bm{A}^{\rm ss}$.

Consider then a reaction $\rho'$ that is maintained out of equilibrium (i.e., $A^{\rm ss}_{\rho'} \neq 0$ {and $J^{\rm ss}_{\rho'} \neq 0$, with the same sign). In any conformal decomposition of $\bm J^{\rm ss}$, at least one EFM containing $\rho'$ must appear.} Let $\bm{e}^*$ denote {one such} EFM, {so that} $\bm{e}^*_{\rho'}\neq 0$ {and $e^*_{\rho'}A^{\rm ss}_{\rho'}>0$.}
We can rewrite Eq.~\eqref{eq:cycle_affinity} to isolate the affinity of $\rho'$:
\begin{equation}\label{eq:start_eq_affinity}
e_{\rho'}^* A_{\rho'}^{\rm ss} =A_{\bm e^*}^Y-\sum_{\rho\neq \rho'}e^*_{\rho}A_{\rho}^{\rm ss}.
\end{equation}
Since $\bm{e}^*$ is conformal to $\bm{A}^{\rm ss}$, all the terms in the summation are positive, so that:
\begin{equation}
    0 < e^*_{\rho'} A_{\rho'}^{\rm ss} <  A_{\bm e^*}^Y.
\end{equation}
This bound relies on {identifying an EFM in a conformal decomposition of the steady-state flux}, which in turn requires knowledge of $\bm{J}^{\rm ss}$ and, consequently, of the entire kinetics. To proceed further and derive a purely thermodynamic bound, we maximize over all EFMs {whose component along $\rho'$ has the same sign as $e^*_{\rho'}$, i.e., over all $\bm{e}$ such that} $e_{\rho'}e^*_{\rho'}>0$. Thus:
\begin{equation}
    0 < {\rm sgn}(e_{\rho'}^*) A_{\rho'}^{\rm ss} <  \frac{A_{\bm e^*}^Y}{|e_{\rho'}^*|} \leq \max_{{\bm e}:
    ~e_{\rho'}e_{\rho'}^*>0} \frac{A_{\bm e}^Y}{|e_{\rho'}|} \;.
    \label{eq:bound2}
\end{equation}
{Since the sign of $e^*_{\rho'}$ depends on the unknown steady-state flux, we distinguish the two possible cases, $e^*_{\rho'}>0$ and $e^*_{\rho'}<0$.} By defining $\mathscr{E}_{\rho'}^{+(-)}$ as the set of all EFMs whose component $e_{\rho'}>0(<0)$, we have:
\begin{eqnarray}
    \label{b1}
    0 < &A_{\rho'}^{\rm ss}& < \max_{\mathscr{E}_{\rho'}^+} \frac{A_{\bm{e}}^Y}{e_{\rho'}} \qquad \textrm{if } e^*_{\rho'} > 0 \\
    \label{b2}
    \min_{\mathscr{E}_{\rho'}^-} \frac{A_{\bm{e}}^Y}{e_{\rho'}} < &A_{\rho'}^{\rm ss}& < 0 \qquad\qquad\;\;\;\ \textrm{if } e^*_{\rho'} < 0
\end{eqnarray}
where we used the fact that, when $e^*_{\rho'}<0$, the right-most side of Eq.~\eqref{eq:bound2} becomes equal to $\max_{\mathscr{E}_{\rho'}^-} (A_{\bm{e}}^Y/|e_{\rho'}|) = \max_{\mathscr{E}_{\rho'}^-} (-A_{\bm{e}}^Y/e_{\rho'}) = - \min_{\mathscr{E}_{\rho'}^-} (A_{\bm{e}}^Y/e_{\rho'})$, by exploiting the negativity of $e_{\rho'}$. {\color{black}Since $\bm e$ and $-\bm e$ represent the same EFM with opposite global orientation and the ratio $A^Y_{\bm e}/e_\rho$ is invariant under this sign reversal, either orientation can be used when taking the extrema over EFMs.} Plugging together Eqs.~\eqref{b1} and \eqref{b2}, for an arbitrary $\rho$, we finally have:
\begin{equation}\label{eq:affinity_bound_1}
\min\left(0,\min_{\mathscr{E}_\rho^-} \frac{A_{\bm e}^Y}{e_{\rho}}\right)\leq A_\rho^{\rm ss}\leq \max\left(0, \max_{\mathscr{E}_\rho^+}  \frac{A_{\bm e}^Y}{e_{\rho}}\right) \;,
\end{equation}
where extremal values are taken over all EFMs containing $\rho$. The distinction between $\mathscr{E}_\rho^+$ and $\mathscr{E}_\rho^-$ only keeps track of the two possible signs of the $\rho$-component, while the ratio $A^Y_{\bm{e}}/e_\rho$ is invariant under global reversal of an EFM. Therefore, the final bound does not require knowing the sign of the steady-state flux or identifying which EFMs are conformal to it, thereby lifting the necessity to specify the entire kinetics. 
Here, we included the equal sign on both inequalities by noting that, at equilibrium, equalities should hold. Hence, although the derivation has been carried out for non-equilibrium systems, it can be readily extended to equilibrium ones. In fact, in a system without non-equilibrium drivings induced by chemostatted species, all cycle affinities vanish ($A_{\bm{e}}^Y = 0$ for all cycles). This condition ensures that, at stationarity, all reaction affinities also vanish ($A_{\rho}^{\rm ss} = 0$ for all reactions), thus the steady state is an equilibrium state and the equalities hold. In the presence of non-equilibrium driving, however, Eq.~\eqref{eq:affinity_bound_1} uncovers the thermodynamic limits of the deviation from equilibrium for each reaction affinity, solely based on global out-of-equilibrium properties. We remark here that the inequality in Eq.~\eqref{eq:affinity_bound_1} turns into an equality also when the presence of non-equilibrium driving gives a zero net flux and, as such, zero affinity. The system effectively behaves as an equilibrium system and equalities are restored as discussed above. Additionally, equalities in Eq.~\eqref{eq:affinity_bound_1} also hold when $\rho$ is a reaction not participating in any cycle since it satisfies detailed balance and, as a consequence, exhibits zero affinity.
A crucial observation is that this formulation bounds local quantities as reaction affinities with global thermodynamic features that do not require knowledge of the EFMs conformal to $\bm{J}^{\rm ss}$. This result holds for any reaction in the CRN.

Furthermore, the affinity of a reaction $\rho$ can be decomposed into contributions from internal and external species, i.e., $A_\rho = A_\rho^{X,\rm ss}+A_\rho^Y$ (see Eq.~\eqref{eq:affinity_decompose}). By moving $A_\rho^Y$ to the right side of Eq.~\eqref{eq:affinity_bound_1}, we obtain an upper bound on the affinity of internal species:
\begin{equation}\label{eq:affinity_bound_2}
    \begin{aligned}
    A_\rho^{X,\rm ss}
    &\leq \max\left(-A_\rho^Y, \max_{\mathscr{E}_\rho^+}  \frac{A_{\bm e}^Y-e_\rho A_\rho^Y}{e_{\rho}}\right)\\
    &=\max_{\Pi^X_{\rho^-}} A_{\bm{\pi}^X_{\rho^-}}^Y
    \end{aligned}
\end{equation}
Here, we define $\Pi^X_{\rho^-}$ as {a set of reaction pathways, inherited from EFMs containing $\rho$,} that convert the internal species $X$ from the right side of reaction $\rho$ to its left side, as indicated by the minus sign. {More precisely, each} element of $\Pi^X_{\rho^-}$ is indicated by $\bm{\pi}^X_{\rho^-}$. This set contains the reaction itself with its affinity, by construction, taken with the minus sign, and all pathways constructed from EFMs where the reaction $\rho$ itself has been eliminated. Indeed, the last equality in Eq.~\eqref{eq:affinity_bound_2} can be easily understood by noting that all EFMs such that $e_\rho>0$, i.e., those in $\mathscr{E}_\rho^+$, convert $X$ from its value on the right to the left side of $\rho$ when $\rho$ itself is eliminated. The affinity of a pathway constructed from $\bm{e} \in \mathscr{E}_\rho^+$ is $A_{\bm{e}}^Y/e_\rho - A^Y_\rho$, where we divided by $e_\rho$ since there might be more conversion of $X$ through $\rho$ in some EFMs. Also, before the elimination of $\rho$, this set of pathways that we named $\Pi^X_{\rho^-}$ contains $\rho$ itself in the backward direction. After elimination, the resulting affinity is simply $-A_\rho^Y$. From these observations, Eq.~\eqref{eq:affinity_bound_2} immediately follows. An analogous procedure yields a lower bound on $A_\rho^{X,\rm ss}$ by taking the negative of the maximum affinity of all pathways in $\Pi^X_{\rho^+}$, i.e., those converting $X$ from the left to the right side of $\rho$.

\subsection*{Derivation for effective reactions in $\mathscr{R}$}

Despite their generality, Eqs.~\eqref{eq:affinity_bound_1} and \eqref{eq:affinity_bound_2} apply only to reaction affinities that are present in the CRN. However, it is possible to surpass this limitation and derive upper and lower thermodynamic bounds for the effective affinity of reactions that are not present in the original CRN, while being in principle compatible with it.

Recalling that the reaction space $\mathscr{R}$ is defined as the space orthogonal to the conservation laws of the original stoichiometric matrix, it contains all possible reactions that are in principle compatible with the CRN under study. We introduce the idea of a ``chemical probe", a test reaction that belongs to the reaction space but not to the original network. The resulting extended stoichiometric matrix is then $\tilde{\bS} = [\bS , \hat{\bS}_{\hat{\rho}}]$, where the newly introduced reaction is represented by the column $\hat{\bS}_{\hat{\rho}}$. We assume that $\hat{\rho}$ proceeds so slowly that its net flux is negligible, i.e., $J_{\hat{\rho}} \simeq 0$. However, it still maintains a finite affinity $A_{\hat{\rho}} = RT \ln[J_{\hat{\rho}}^+ / J_{\hat{\rho}}^-]$, which depends on the chemical potentials of the species involved in the reaction. To determine the affinity associated with this test reaction $\hat{\rho}$, we consider one conformal EFM containing $\hat{\rho}$, $\bm{e}^{(F)}$. For this, according to the previous relationship, we have:
\begin{equation}
\label{probe1}
    {\rm sgn}\left(e^{(F)}_{\hat\rho}\right)A_{\hat\rho}^{\rm ss} < \max_{\mathscr{E}^+_{\hat\rho}} \frac{A^Y_{\bm {e}}}{|e_{\hat\rho}|} \;.
\end{equation}
{Since the reaction $\hat{\rho}$ induced by the chemical probe belongs to the reaction space of the original CRN, adding it closes at least one cycle involving the probe and a pathway $\bm{\pi}^{(F)}$ of the original network, $\bm{e}^{(F)} = \bm{\pi}^{(F)} + \bm e_{\hat{\rho}}$, where $\bm e_{\hat{\rho}}$ is the unit vector along the probe reaction. Moreover, because the probe is introduced with infinitesimal flux, the steady-state affinities of the original reactions are unchanged. Besides the EFM $\bm e^{(F)}$ that contains $\hat{\rho}$ and is aligned with both $A^{\rm ss}_{\hat{\rho}}$ and the steady-state affinities along $\bm{\pi}^{(F)}$, one can therefore construct a second flux mode by adding to the probe-containing cycle a flux mode of the original CRN that passes through the same pathway $\bm{\pi}^{(F)}$ with opposite orientation and is closed by a different pathway. This operation leaves the probe orientation unchanged, cancels the contribution of $\bm{\pi}^{(F)}$, and yields a flux mode whose components on the original reactions it contains have signs opposite to the corresponding steady-state affinities. A conformal decomposition of this flux mode then contains an EFM with the same sign structure, which we denote by $\bm e^{(B)}$.}
Therefore, we have that $e_{\hat{\rho}}^{(B)} A^{\rm ss}_{\hat{\rho}} > 0$ and $e_{\rho}^{(B)} A^{\rm ss}_{\rho} \leq 0$ for $\rho \neq \hat{\rho}$.
\begin{figure}[t]
    \includegraphics[width=1\columnwidth]{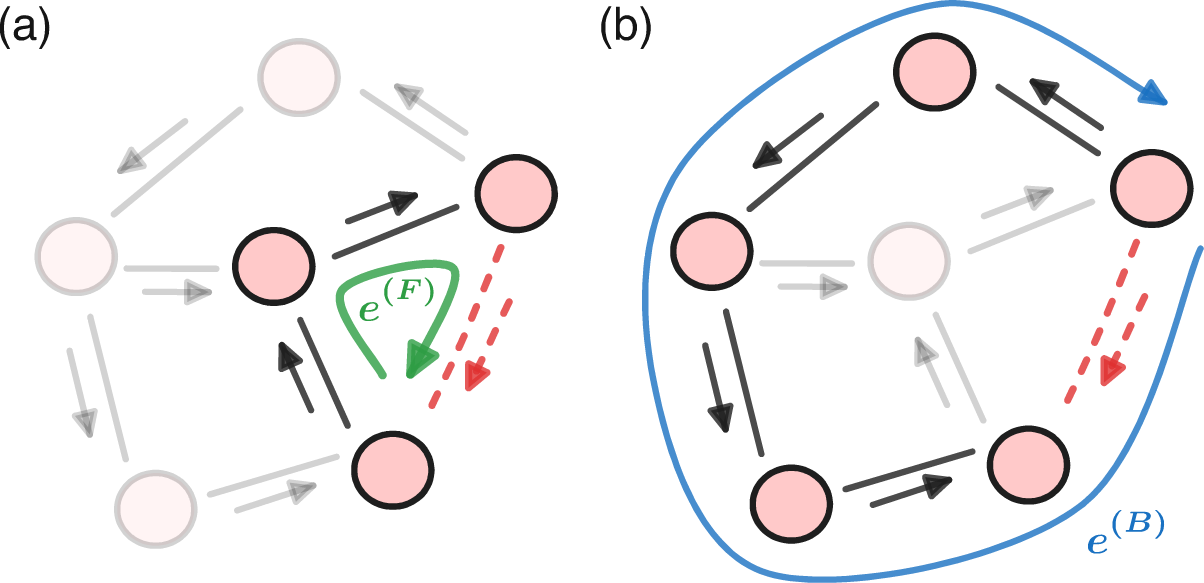}
    \caption{\textbf{Example of a chemical probe.} (a) A linear CRN is shown with all reactions indicated by black (and gray) lines and the probe indicated by a dashed red line. Arrows denote stationary reaction affinities. The stationary flux associated with the probe vanishes, but its affinity remains finite. In green, we show the EFM conformal to the affinities of the modified network (i.e., original CRN plus the probe), $\bm{e}^{(F)}$. We highlighted nodes and edges involved in this EFM. (b) The EFM of the modified network aligned with the probe affinity, but anti-aligned with the steady-state affinities of the original CRN, $\bm{e}^{(B)}$. Involved edges and nodes are highlighted.}
    \label{fig:chem_probe}
\end{figure}
In Fig.~\ref{fig:chem_probe}, we illustrate this idea with a simple CRN to build direct intuition. We also remark that anti-alignment holds for all edges participating in a cycle in the original CRN. In all other cases, edges not belonging to any cycle can just be ignored as they do not contribute to the cycle affinity. By implementing these considerations in Eq.~\eqref{eq:start_eq_affinity}, we have:
\begin{equation}
\label{probe2}
    {\rm sgn}\left(e^{(B)}_{\hat\rho}\right)A_{\hat\rho}^{\rm ss} > \frac{A^Y_{\bm{e}^{(B)}}}{|e_{\hat\rho}^{(B)}|} \geq \min_{\mathscr{E}^{+}_{\hat\rho}} \frac{A^Y_{\bm{e}}}{|e_{\hat\rho}|} \;,
\end{equation}
by using the anti-alignment along all reactions but $\hat\rho$. We can arbitrarily fix the sign of $e^{(F)}_{\hat\rho}$ to be positive and, by construction, the sign of $e^{(B)}_{\hat\rho}$ will be positive as well. Thus, by putting together Eqs.~\eqref{probe1} and \eqref{probe2}, we have:
\begin{equation}
\label{eq:probe_affinity_bound}
    \min_{\mathscr{E}_{\hat\rho}^-} \frac{A^Y_{\bm{e}}}{e_{\hat\rho}} \leq A^{\rm ss}_{\hat\rho} \leq \max_{\mathscr{E}_{\hat\rho}^+} \frac{A^Y_{\bm{e}}}{e_{\hat\rho}} \;,
\end{equation}
where we used that equalities describe the equilibrium condition, as before, and that $\min_{\mathscr{E}_{\hat\rho}^+} A^Y_{\bm{e}}/e_{\hat\rho} = \min_{\mathscr{E}_{\hat\rho}^-} A^Y_{\bm{e}}/e_{\hat\rho}$, since both affinities and $e_{\hat\rho}$ change sign. It is immediate to check that, if ${\rm sgn}(e^{(F)}_{\hat\rho}) = {\rm sgn}(e^{(B)}_{\hat\rho}) < 0$, this sandwich inequality still holds. Crucially, Eq.~\eqref{eq:probe_affinity_bound} provides a tighter bound with respect to Eq.~\eqref{eq:affinity_bound_1}, since we used the additional information characterizing the chemical probe. This sandwich inequality holds for any reaction belonging to the reaction space $\mathscr{R}$ and determines the thermodynamic limits of any effective affinity, given an energy budget determining non-equilibrium conditions. Therefore, Eqs.~\eqref{eq:affinity_bound_1} and \eqref{eq:probe_affinity_bound} constitute the first main result of this study.

By splitting the affinity into the contribution given by internal and external species, we can rewrite the upper bound as follows:
\begin{equation}
\label{eq:probe_affinity_bound2}
    A^{X,\rm{ss}}_{\hat\rho} \leq \max_{\mathscr{E^+_{\hat\rho}}} \frac{A^Y_{\bm{e}} - e_{\hat\rho} A^Y_{\hat\rho}}{e_{\hat\rho}} = \max_{\Pi^X_{\hat\rho^-}} A^Y_{\pi_{\hat\rho^-}^X}
\end{equation}
and the same holds for the lower bound. There is one main difference between this expression, valid for the affinity of a reaction not present in the original CRN, and the one presented in Eq.~\eqref{eq:affinity_bound_2}. In the first case, the term given by the direct reaction $\hat\rho$ does not appear and, as such, the pathways converting $X$ from the right to the left of $\hat\rho$ have to be considered in the original CRN. This property is perfectly consistent with the notation here introduced and with the fact that this chemical probe is only a mathematical tool to investigate the system devoid of any physical information. As a final comment, such a chemical probe can be considered analogous to a voltmeter in an electric circuit. Both measure potential differences with minimal disturbance: the probe is introduced gradually, resulting in an infinitely slow edge of the network, while an ideal voltmeter has infinite resistance, leading to infinitesimal flux. As such, the probe uses its vanishing chemical flux to gauge the chemical potential difference between network nodes, similarly to how a voltmeter measures voltage. As we will see in the next section, this approach will be crucial to determine the whole thermodynamic space of a CRN.

\subsection{Thermodynamic Space of CRNs - Accessible region for species concentrations}

From the bounds on reaction affinities, we can derive upper and lower bounds on species concentrations. These limits determine the accessible region for each chemical species given the CRN's energy budget. We name this region ``thermodynamic space''. The same concept has been already introduced in \cite{liang2024Thermodynamic}, where it has been found for linear and catalytic CRNs. Also, the idea of a space which is thermodynamically accessible given the external non-equilibrium constraints is of fundamental importance in various contexts, such as metabolic networks \cite{famili2003Convex}, signal amplification \cite{arunachalam2024Informationa}, and self-assembly \cite{ragazzon2018energy}. Once again, the result of this part is summarized in Sec.~\ref{sec:outline}~B, while its derivation is presented below.\\

\subsection*{Derivation of the Thermodynamic Space}

{Although stationary concentrations are not determined by thermodynamics alone, since they generally depend on the kinetic properties of a CRN, they enter the thermodynamic description through the chemical potentials.} We recall that affinities can also be expressed by means of chemical potentials. By specifying Eq.~\eqref{eq:affinity_vector} for $A^{X, \rm{ss}}_\rho$ and $A^Y_{\bm{\pi}_{\rho^-}^X}$, we have:
\begin{equation}\label{eq:affinity_free_energy}
    \begin{aligned}
    &A_\rho^{X, \rm{ss}} 
    = -\Delta_\rho G_X^{\circ} - RT\sum_i\bS_{i\rho}^X\ln x_i^{\rm ss},\\
    &A_{\bm{\pi}^X_{\rho^-}}^Y 
    = -\Delta_{\bm{\pi}^X_{\rho^-}} G_Y \;,
    \end{aligned}
\end{equation}
where $\Delta_\rho G_X^{\circ}$ indicates the standard Gibbs free-energy change of $X$ along $\rho$, and $\Delta_{\bm{\pi}^X_{\rho^-}} G_Y$ the Gibbs free-energy change of $Y$ along the reaction pathway $\bm{\pi}^X_{\rho^-}$.
By combining Eqs.~\eqref{eq:affinity_bound_2} and \eqref{eq:affinity_free_energy}, and employing the duality of affinity, i.e., $A^Y_{\bm{\pi}^X_{\rho^-}} = - A^Y_{\bm{\pi}^X_{\rho^+}}$ for any $\bm{\pi}^X_{\rho^-}$ so that:
\begin{equation}
    \max_{\Pi^X_{\rho^-}} A^Y_{\bm{\pi}_{\rho^-}^X} = - \min_{\Pi^X_{\rho^+}} A^Y_{\bm{\pi}_{\rho^+}^X} \;,
\end{equation}
we obtain bounds on the internal concentrations involved in the reaction $\rho$:
\begin{equation}\label{eq:eq_constant_bound_0}
    \begin{aligned}
        \prod_i (x_i^{\rm ss})^{\bS_{i\rho}^X}
        &\geq \min_{\Pi^X_{\rho^+}} \underbrace{ \exp\left[-\frac{\Delta_\rho G_X^{\circ}+\Delta_{\bm{\pi}^X_{\rho^+}} G_Y}{RT}\right]}_{K_{\bm{\pi}^X_{\rho^+}}} \equiv K_{\bm{\pi}^X_{\rho^+}}^{\rm min}
    \end{aligned}
\end{equation}
Here, $K_{\bm{\pi}^X_{\rho^+}}$ is the effective equilibrium constant for a reaction pathway converting chemical species from the left to the right side of reaction $\rho$. Indeed, if the CRN only had one pathway, say $\bm{\pi}^X_{\rho^+}$, the system would reach equilibrium and the product of concentrations of species involved in $\rho$ with stoichiometric exponents, $\prod_i (x_i^{\rm ss})^{\bS_{i\rho}^X}$, would be exactly equal to $K_{\bm{\pi}^X_{\rho^+}}$.
As indicated in Eq.~\eqref{eq:eq_constant_bound_0}, the concentrations are lower bounded by the equilibrium constant minimized over all these pathways. Following the same procedure, we can derive also an upper bound, resulting in the following range of accessible concentrations for all species involved in $\rho$:
\begin{equation}\label{eq:eq_constant_bound}
    K_{\bm{\pi}^X_{\rho^+}}^{\rm min}\leq \prod_i (x_i^{\rm ss})^{\bS_{i\rho}^X}\leq  K_{\bm{\pi}^X_{\rho^+}}^{\rm max}
\end{equation}
The bounds in Eq.~\eqref{eq:eq_constant_bound} have been derived for the concentration of species connected by a reaction $\rho$. However, starting from Eq.~\eqref{eq:probe_affinity_bound2}, we can see that analogous bounds hold for any species connected by a fictitious reaction $\hat\rho$ which is compatible with the conservation laws, i.e., it belongs to the reaction space of the CRN. These bounds are identical to Eq.~\eqref{eq:eq_constant_bound} by replacing $\rho$ with $\hat\rho$ and the pathways have to be constructed from the original CRN (as the chemical probe never existed), consistently with the physical content of the system. Thus, Eq.~\eqref{eq:eq_constant_bound} constitutes the second main result of this study.

{From a computational perspective, determining these thermodynamic bounds does not necessarily require full EFM enumeration that, as discussed above and shown in Appendix \ref{sec:EFM_enumeration}, is feasible for small to medium CRNs. In fact, only the pathways involving driven reactions have to be identified, as those that extremize the thermodynamic contributions in Eq.~\eqref{eq:eq_constant_bound} are needed to determine the Thermodynamic Space. Thus, the present formulation naturally points to a targeted EFM search problem containing prescribed reactions. Existing algorithms in this direction have been recently proposed \cite{david2014computing}, while the development of refined targeted approaches that mitigate the combinatorial explosion associated with full EFM determination remains an active area of research \cite{klamt2017elementary,ullah2020towards}, and lies beyond the scope of this work.}

{We remark that each bound in Eq.~\eqref{eq:eq_constant_bound} can be saturated in unusual pseudo-equilibrium conditions with vanishing fluxes that are attained by a CRN kinetically dominated by one single pathway. Such a pathway is the one determining the extremal contribution to the free-energy difference, i.e., that leading to Eq.~\eqref{eq:eq_constant_bound_0} for the lower bound. The saturation conditions are analogous to those discussed in \cite{liang2024Thermodynamic} and \cite{qureshi2026Thermodynamic} for linear systems.}

Let us investigate the difference between equilibrium and non-equilibrium conditions and the information that the thermodynamic space can provide on the system. At thermodynamic equilibrium, all driving forces vanish and, as such, the equilibrium constant becomes identical for all pathways and equal to the standard Gibbs free-energy difference of internal species, as expected. 
Consequently, Eq.~\eqref{eq:eq_constant_bound} becomes an equality on both sides, thus the stationary value of all concentrations converges to a unique one dictated by their free energies and the conservation laws of the CRN. In contrast, out-of-equilibrium CRNs exhibit a range of feasible stationary concentrations. This constitutes a subspace of the whole phase-space of species concentrations, and it depends on the global non-equilibrium thermodynamic properties of the network. In general, a non-zero thermodynamic space enables complex dynamics and is a prerequisite for the emergence of complex behaviors such as free-energy transduction, bi-stability, pattern formation, and kinetically responsive states. The thermodynamic space, encoded in the bounds of Eq.~\eqref{eq:eq_constant_bound}, quantifies the intimate relationship between non-equilibrium conditions and emergent complexity in any CRN, therefore providing a measure of potential functional diversity and efficiency of complex biochemical processes.
\dmb{\subsection{Thermodynamic Space to test candidate reaction topologies\label{sec:test_topologies}}}

\begin{figure}[!htb]
    \centering
    \includegraphics[width=1\columnwidth]{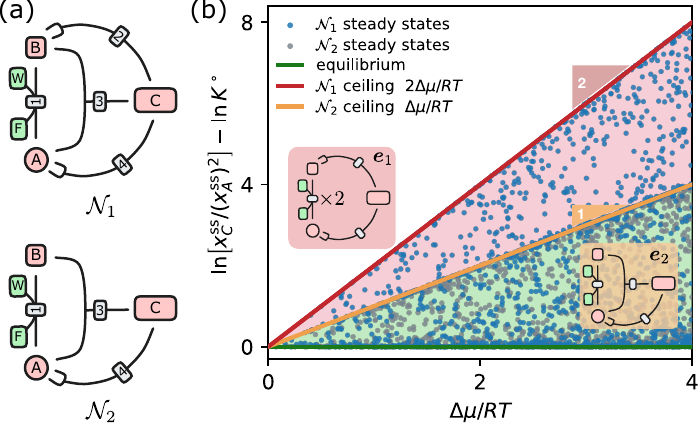}
    \caption{\textbf{Testing reaction topologies with the Thermodynamic Space.} (a) Two candidate topologies for the network of Fig.~\ref{fig:CRN_intro}: $\mathcal{N}_1$ contains reactions $1$, $2$, $3$, $4$, while $\mathcal{N}_2$ removes the reaction $2$ ($2B\rightleftharpoons C$). (b) Accessible range of the dimer amplification $x_C^{\rm ss}/(x_A^{\rm ss})^2$, measured relative to its equilibrium value $K^\circ$, as a function of the driving $\Delta\mu$. The equilibrium value (green line) is the shared lower bound; the mode $\bm{e}_2$ experiences only once the driven reaction and sets the ceiling at slope $1$. This defines the upper bound of the Thermodynamic Space of $\mathcal{N}_2$ (green shaded area); Conversely, the mode $\bm{e}_1$, accessible only when reaction 2 is present, uses twice the driven reaction and raises the upper bound to slope $2$. This sets the upper bound of the Thermodynamic Space of $\mathcal{N}_1$. The region between the two ceilings (red shaded area) is the inference window for reaction 2. Dots are stationary states from simulations with random kinetic rates and total concentrations: those of $\mathcal{N}_2$ (grey) stay below the slope-$1$ ceiling, while those of $\mathcal{N}_1$ (blue) populate the window, so a measured value lying in the inference window cannot arise from $\mathcal{N}_2$ and reveals the existence of $2B\rightleftharpoons C$ reaction.}
    \label{fig:topology_inference}
\end{figure}

{

The bounds derived above require the specification of the CRN topology. However, they can be turned around into a tool that interrogates the reaction network itself. Because the Thermodynamic Space defined by Eq.~\eqref{eq:eq_constant_bound} contains every stationary state reachable by any kinetics compatible with a given topology, a measured concentration lying outside its limits cannot be explained by tuning rate constants. Such a result would falsify the whole class of kinetics consistent with the proposed topology. Therefore, the Thermodynamic Space is a kinetics-free consistency test for a candidate set of reactions, and, as we now show for hypergraphs, a violation can even point to the specific reaction that is missing.

Consider the network in Fig.~\ref{fig:CRN_intro}b (also reported in Fig.~\ref{fig:topology_inference}a for completeness and indicated as $\mathcal{N}_1$), a dimer can be assembled through two active monomers ($2B\rightleftharpoons C$), one active and one passive ($A+B\rightleftharpoons C$), or two passive monomers ($2A\rightleftharpoons C$). The dimer amplification $\ln [x_C^{\rm ss}/(x_A^{\rm ss})^2]-\ln K^{\circ}$, {where $K^\circ=e^{-(\mu_C^\circ-2\mu_A^\circ)/RT}$}, is upper bounded by the elementary flux mode that contains the fuel-driven reaction $F+A\rightleftharpoons B+W$. In particular, the mode $\bm{e}_1=[2,1,0,-1]^T$, which activates both monomers before assembly, experiences twice the driven reaction, thereby setting the ceiling as
\begin{equation}\label{eq:topo_bound_full}
     0\le\ln\frac{x_C^{\rm ss}}{(x_A^{\rm ss})^2}-\ln K^\circ \le 2\Delta\mu/RT \;.
\end{equation}
In Fig.~\ref{fig:topology_inference}b, the upper bound and its corresponding mode $\bm{e}_1$ are shown in red.

Imagine now removing reaction 2, $2B\rightleftharpoons C$. The corresponding network is shown in Fig.~\ref{fig:topology_inference}a and indicated as $\mathcal{N}_2$. This elimination would destroy the reaction mode $\bm{e}_1$ and, as a consequence, the remaining mode $\bm{e}_2=[1,0,1,-1]^T$ would set the upper bound to dimer amplification. However, in $\bm{e}_2$, the driven reaction only appears once, thus lowering the ceiling to
\begin{equation}\label{eq:topo_bound_reduced}
    0\le \ln\frac{x_C^{\rm ss}}{(x_A^{\rm ss})^2} -\ln K^\circ\le \Delta\mu/RT \;.
\end{equation}
In Fig.~\ref{fig:topology_inference}b, this upper bound and its corresponding mode $\bm{e}_2$ are shown in yellow.

Therefore, in this simple example, a single added hyperedge doubles the exponent of the attainable amplification, leading to an inference window $[\Delta\mu/RT,2\Delta\mu /RT]$ that cannot be captured without the active-monomer assembly reaction, and reveals the presence of this hidden reaction with no fit to rate constants.

This argument mainly works as a falsification test for candidate topologies and can be extended to all cases in which a driven reaction can be used more than once by adding specific hyperedges. Clearly, the re-use of a driven reaction has no analogue in linear networks and is a genuine signature of the underlying hypergraph structure. This inference method is one-sided and robust: exceeding a ceiling falsifies a topology, whereas a candidate missing reaction that does not re-use the driving leaves the bounds unchanged. Our approach therefore identifies
the class of missing pathways that can be spotted in principle, thereby complementing the data-driven inference of driving forces discussed in Section \ref{sec:data}, which instead focuses on inferring the energetics of a given CRN topology.}\\

\subsection{Connections to previous results\label{sec:connection_to_previous}}
In this work, we derived two main results: the thermodynamic constraints on reaction affinity at stationary state in terms of cycle affinities, i.e., Eqs.~\eqref{eq:affinity_bound_1} and \eqref{eq:probe_affinity_bound}, and the bounds on species concentrations in terms of (pseudo-)equilibrium constants, i.e., Eq.~\eqref{eq:eq_constant_bound}, that define the thermodynamic space of the CRN. Similar results have been previously established for special cases of CRNs such as linear and catalytic CRNs. Therefore, our results generalize and extend these findings to the challenging case of a generic open chemical system, introducing the concept of Thermodynamic Space and providing a method to find it for any CRN.

In linear and catalytic CRNs, entries of each column of the stoichiometric matrix are restricted to have only one $+1$ and one $-1$ for internal species
, effectively making the CRN a graph rather than a hypergraph. For such graph networks, a simple conservation law applies $\bm{r} = \bm{1}$, i.e., $\sum_i x_i = \mathrm{constant}$. Moreover, the matrix-tree theorem allows the representation of steady-state solutions through spanning trees, facilitating the derivation of thermodynamic bounds on steady-state observables. Previous results, primarily discussed in the context of stochastic thermodynamics, expressed bounds in terms of probability ratios. These probabilities can be interpreted as normalized concentrations of internal species with respect to the conservation law, i.e., $p_i = x_i / \sum_i x_i$. This formulation allowed for a direct connection between deterministic and stochastic descriptions of the system:
\begin{equation}
    K_{{\bm{\pi}}_{X_i\to X_{i'}}}^{\rm min}\leq \frac{p_{i'}^{\rm ss}}{p_i^{\rm ss}}\leq K_{{\bm{\pi}}_{X_i\to X_{i'}}}^{\rm max},
\end{equation}
which can directly come from Eq.~\eqref{eq:eq_constant_bound} by choosing a chemical probe $X_i\rightleftharpoons X_{i'}$ such that $\bS_{i\hat{\rho}}=-1$ and $\bS_{i'\hat{\rho}}=1$. An equivalent form of the above result was first discovered by Maes and Neto\v{c}n\'{y} for linear networks \cite{maes2013Heat}, in which upper and lower bounds were expressed using the heat dissipation through minimal and maximal dissipation pathways, i.e. rewriting the (pseudo-)equilibrium constants in terms of free energy changes. Lin has derived a similar result by drawing an analogy between electric circuits and linear CRNs \cite{lin2020Circuit}. Subsequently, our previous work demonstrated that these bounds also hold for non-linear catalytic networks \cite{liang2024Thermodynamic}, where catalytic effects do not change the entries of the stoichiometric matrix and maintain the CRN as an effective graph instead of a hypergraph. Additionally, Liang and Pigolotti used a trajectory-based approach to prove these bounds for linear networks and extended their analysis to constrain steady-state flux asymmetries \cite{liang2023Thermodynamic}:
\begin{equation}
    \left|\ln\frac{J_\rho^+}{J_\rho^-}\right|\leq \max_{\bm{e}\ni\rho}\frac{A_{\bm{e}}^Y}{RT}.
\end{equation}
This result can be seen as a special case of Eq.~\eqref{eq:affinity_bound_1} for linear CRNs, expressing the ratio of forward and backward semi-fluxes using the LDB condition, Eq.~\eqref{eq:LDB}.

Ultimately, we remark that the result presented here encompasses all previous work, generalizing their findings (both on affinities and species concentrations) to general CRNs. Such extension constitutes the first mathematical derivation of the range of operations of chemical systems, and is of crucial importance. In fact, many central biological processes, from metabolism and signaling to self-assembly \cite{palsson2015Systems, zoller2022Eukaryotic, ragazzon2018energy}, as well as emergent complex phenomena like multistability and pattern formation \cite{vellela2009Stochastic, cao2015Freeenergy, liang2024Thermodynamic}, fundamentally rely on multi-molecular interactions inherent to such generic networks, thus requiring a framework beyond simpler models. Therefore, our results can unravel general properties of a variety of different systems. They can reveal intrinsic potentialities and limitations due to non-equilibrium conditions, and lead to universal answers in multiple fields, ranging from proofreading to the origin of life, where hitherto only solutions for specific systems have been obtained.


\section{Applications\label{sec:application}}

\subsection{Schl\"ogl model for bistability\label{sec:schlogl_model}}
Out-of-equilibrium open CRNs may exhibit emergent complex phenomena, such as multistability, where the reaction dynamics admits multiple stable fixed points. This non-equilibrium multistability is fundamental to various biological functions, including gene regulation, signal transduction, and cell fate decisions \cite{tiwari2011Bistable,qian2005Nonequilibrium,wang2009Bistable}. Yet, understanding its quantitative constraints imposed by thermodynamic driving remains a challenging question. The Schl\"ogl model, a minimal model for multistability, can sustain two stable stationary states within specific parameter ranges \cite{vellela2009Stochastic,schlogl1972Chemical}. Analyzing this feature in such a paradigmatic model is crucial to elucidate the role of non-equilibrium thermodynamic driving forces in constraining and shaping the emergent bistable behaviour of chemical reaction systems.

\begin{figure}[t]
    \includegraphics[width=1\columnwidth]{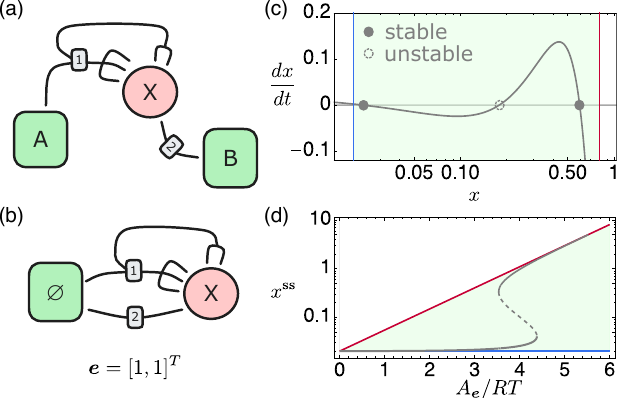}
    \caption{\textbf{Thermodynamic space of the Schl\"ogl model.} (a) Reaction hypergraph of the Schl\"ogl model with an internal species $X$ and two chemostatted external species $A$ and $B$. (b) The EFM of the Schl\"ogl model, in which the internal species $X$ is created from the environment (indicated as $\emptyset$) and then degraded into it through reactions 1 and 2. (c) Nullcline of the model.
    Fixed points (zeros of the derivative) are constrained within the thermodynamic space (green-shaded region). (d) Bifurcation diagram with cycle affinity $A_{\bm{e}} = \mu_A - \mu_B$ as the control parameter. Stable (solid lines) and unstable (dashed lines) solutions are bounded by the thermodynamic space (green-shaded region). Blue and red lines in (c) and (d) represent the lower and upper bound of the thermodynamic space, respectively, as derived from Eq.~\eqref{eq:con_bound_Schlogl}. Numerical results in (c)-(d) were obtained with $k_1^{\pm}=8$, $k_2^{\pm} = 1$  and $b = 0.02 $, while the driving force was varied by adjusting the chemostatted external concentration $a$. In (c), $a = 0.8$.}
    \label{fig:Schogl}
\end{figure}

The Schl\"ogl model is defined by the following reactions and stoichiometric matrix:
\begin{equation}\label{eq:Schlogl_model}
    \begin{aligned}
       \rboxed{1}\ &2 X + A\xrightleftharpoons[k_1^-]{k_1^+} 3X\\
       \rboxed{2}\ &X\xrightleftharpoons[k_2^-]{k_2^+} B
    \end{aligned}\quad
        \bS   
      =\left[
  \begin{array}{ccc}
    1&-1 \\
    \hline 
    -1 & 0 \\
    0 & 1 \\
  \end{array}
\right]
  \begin{array}{c}
    X \\
    \hline 
    A\\
    B\\
  \end{array}
\end{equation}
where $X$ is an internal species, and $A$ and $B$ are two chemostatted external species. These reactions can be represented by the reaction hypergraph shown in Fig.~\ref{fig:Schogl}(a). 
The network contains a single EFM, $\bm{e} = [1,1]$, represented in Fig.~\ref{fig:Schogl}(b). Along this EFM, one external species $A$ is converted into an external species $B$, with an associated free-energy change $\Delta_{\bm{e}}G =  \mu_B - \mu_A = \mu_B^\circ - \mu_A^\circ + RT\ln(b/a)$. Consequently, its cycle affinity is $A_{\bm{e}}= -\Delta_{\bm{e}}G$. Applying Eq.~\eqref{eq:affinity_bound_1}, we derive bounds on the affinities of both reactions for internal species:
\begin{equation}
    \begin{aligned}
    0\leq A_{\rho = 1}^{\rm ss}\leq \mu_A-\mu_B,\quad
    0\leq A_{\rho = 2}^{\rm ss}\leq \mu_A-\mu_B.
    \label{eq:bound_Sch}
    \end{aligned}
\end{equation}
To constrain the concentrations of internal species, we apply Eq.~\eqref{eq:eq_constant_bound} to all species involved in the first reaction. In this particular case, the same result would also be obtained by considering the second reaction, since we can only bound $X$ as internal species. Indeed, also the bounds on reaction affinities coincide (see Eq.~\eqref{eq:bound_Sch}).
We first noticed that a net production of the internal species, $\emptyset \to X$, can be achieved through two pathways: $\bm{\pi}_{\emptyset\to X}^{(1)} = [1,0]$ (first reaction) and $\bm{\pi}_{\emptyset\to X}^{(2)} = [0,-1]$ (reverse of second reaction). We then identify the minimum and maximum equilibrium constants considering these pathways
and obtain the following bounds on the steady-state concentration of the internal species $X$:
\begin{equation}\label{eq:con_bound_Schlogl}
   \begin{aligned}
    \exp\left[-\frac{\mu_X^{\circ}-\mu_B}{RT}\right]\leq x^{\rm ss}\leq \exp\left[{-\frac{\mu_X^{\circ}-\mu_A}{RT}}\right].
   \end{aligned}
\end{equation}
The range of steady-state concentration constrains the fixed points of the reaction dynamics. In Fig.~\ref{fig:Schogl}(c), we show that this region contains both stable and unstable fixed points. Additionally, in Fig.~\ref{fig:Schogl}(d) we bound the whole bifurcation diagram for different values of the cycle affinity $A_{\bm{e}}=\mu_A-\mu_B$. The lower and upper bounds of Eq.~\eqref{eq:con_bound_Schlogl}, indicated by blue and red lines in Fig.~\ref{fig:Schogl}, can be asymptotically approached in the timescale separation limit. In fact, the system reaches the upper bound when the first reaction proceeds much faster than the second, allowing it to reach a quasi-equilibrium state. Conversely, the lower bound is approached when the second reaction is significantly faster, dominating the system dynamics. This timescale separation principle provides insight into how kinetic parameters can drive the system towards different extremes within its thermodynamically allowed range, highlighting the interplay between thermodynamic constraints and kinetic factors in determining steady-state concentrations.

The nontrivial bistable behavior of this reaction network emerges from the high-order autocatalytic nature of the first reaction. This autocatalysis modulates reaction rates based on the concentration of species $X$, leading to two stable fixed points approaching the upper and lower bounds of the thermodynamic space defined by Eq.~\eqref{eq:con_bound_Schlogl}. These bounds effectively constrain the range of stable fixed points. In equilibrium systems, where $\Delta_{\bm{e}}G = \mu_B - \mu_A = 0$, the lower and upper bounds coincide, collapsing the thermodynamic space and precluding bistable solutions. This collapse elucidates the necessity of non-equilibrium driving forces for the emergence of complex behaviors in chemical reaction networks (CRNs), even for simple models.

\begin{figure*}[t]
    \includegraphics[width=1.9\columnwidth]{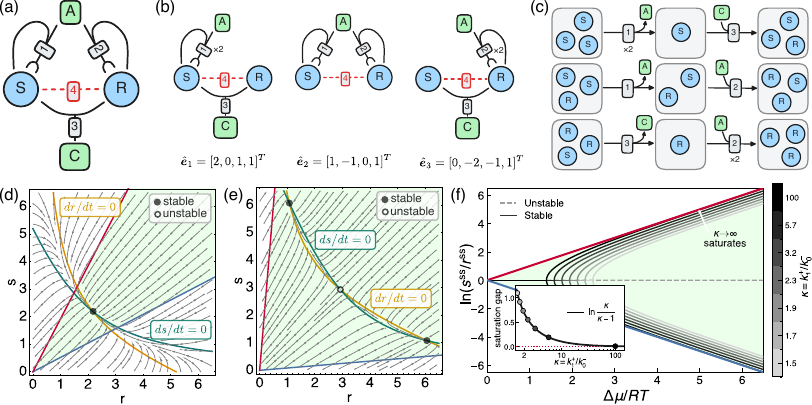} 
    \caption{\textbf{Thermodynamic analysis of chiral symmetry breaking.} (a) Thermodynamically consistent Frank model composed of chiral species $R, S$, achiral precursor $A$, and dimer $C$. $A, C$ (green squares) are chemostatted. The dashed red edge indicate the chemical probe.
    (b) Key Elementary Flux Modes (EFMs) $\hat{\bm{e}}_1, \hat{\bm{e}}_2, \hat{\bm{e}}_3$ involving the probe shown with a vector representation.
    (c) Molecular transformation pathways, derived from the EFMs in (b), that, as a net effect, convert the species from the left to the right side of the probe ($S\to R$)
    (d, e) Phase portraits in the $r-s$ concentration plane, displaying nullclines ($dr/dt=0$, orange; $ds/dt=0$, blue), fixed points (stable: $\bullet$; unstable: $\circ$), and system trajectories (streamlines). The green shaded region is the thermodynamically accessible space defined by Eq.~\eqref{eq:eq_constant_bound}. In (d), $\Delta \mu = 0.7RT$, while $\Delta \mu = 2.5RT$ in (e).
    (f) Pitchfork bifurcation diagram for chiral symmetry breaking: Log-ratio of enantiomer concentrations, $\ln(s^{ss}/r^{ss})$, versus the thermodynamic driving force $\Delta\mu = 2\mu_A - \mu_C$. Solid lines represent stable fixed points; the dashed line indicates the unstable achiral state. The green shaded region, bounded by red and blue lines (theoretical limits at $\ln(s/r) = \pm \Delta\mu/RT$), illustrates the thermodynamic bounds on chiral imbalance.  The stable chiral branches are shown for increasing timescale separation $\kappa=k_1^+/k_0^-$ (gray scale). Making the dimerization fast at fixed $k_1^+/k_1^-$ (hence fixed $\Delta\mu$) holds $R+S\rightleftharpoons C$ at pseudo-equilibrium, so that each branch saturates the bound $\pm\Delta\mu/RT$, reducing the residual gap $\Delta\mu/RT-\ln(r^{\rm ss}/s^{\rm ss}) \approx \ln[\kappa/(\kappa-1)]$ to $0$ as $\kappa \to \infty$ (inset). For each curve, we set $k_0^+=1$, $k_0^-=0.7$, $k_1^-/k_1^+=8.5$, $c=0.5$ and change $a$ to tune $\Delta\mu$.}
    \label{fig:chiral_breaking_fig3}
\end{figure*}

While we present a minimal model for multistability, our thermodynamic constraints apply to arbitrary CRNs and can be potentially extended to complex biochemical processes in cellular systems, illuminating the onset of bistable modes in gene regulation or other biological functions whose robustness resides in the presence of multistability. Moreover, since our results do not require any detailed kinetic information, it provides a tool for understanding thermodynamic constraints that underscore the emergence of complex behaviors in chemical and biological systems within physical limits.

\subsection{Chiral symmetry breaking}

Chiral symmetry breaking (CSB) is fundamental for understanding the origin of life, wherein biomolecular homochirality is thought to arise from spontaneous symmetry breaking mechanisms. While dynamical models rely on autocatalysis and recycling to explain CSB emergence \cite{saito2013Colloquium, plasson2004Recycling, frank1953Spontaneous}, they crucially depend on out-of-equilibrium conditions to differentiate enantiomers of identical energy \cite{blackmond2020Autocatalytic}. Although the necessity of dissipation is recognized, a quantitative understanding of the relationship between the strength of non-equilibrium driving and the degree of CSB has remained elusive. Here, we apply our thermodynamic framework to elucidate this connection.

We analyze a thermodynamically consistent, reversible version of the Frank model \cite{frank1953Spontaneous}, a minimal system known to exhibit CSB. The model describes two chiral isomers, $R$ and $S$, that are produced autocatalytically from an achiral precursor $A$, and subsequently undergo dimerization to form a complex $C$ (see Fig.~\ref{fig:chiral_breaking_fig3}a). All reactions are considered reversible and adhere to local detailed balance (see Eq.~\eqref{eq:LDB}). The species $A$ and $C$ are chemostatted, with their chemical potentials establishing the non-equilibrium driving. The reactions and stoichiometric matrix are given by
\begin{equation}
\label{eq:frank_reaction}
\begin{aligned}
    \begin{aligned}
      \rboxed{1}\  S+A&\xrightleftharpoons[k_0^-]{k_0^+} 2S,\\
      \rboxed{2}\  R+A&\xrightleftharpoons[k_0^-]{k_0^+} 2R,\\
      \rboxed{3}\  R+S&\xrightleftharpoons[k_1^-]{k_1^+} C.
    \end{aligned}\quad
        \bS   
      =\left[
  \begin{array}{ccccc}
     1 & 0 & -1  \\
     0 & 1 & -1  \\
    \hline 
    -1 &-1 &0  \\
    0 & 0 & 1 \\
  \end{array}
\right]
  \begin{array}{c}
    S \\
    R\\
    \hline 
    A \\
    C
  \end{array}
\end{aligned}
\end{equation}
The system's elementary flux mode (EFM) is $\bm{e}=[1,1,1]^T$, corresponding to the net conversion $2A \rightarrow C$. The associated cycle affinity, $A_{\bm{e}}  = \Delta\mu=2\mu_A - \mu_C$, quantifies the thermodynamic driving force.

To evaluate the chiral imbalance at steady state --- quantified by the difference between chemical potentials of chiral species, $\mu^{\rm ss}_R - \mu^{\rm ss}_S$ ---  we employ the chemical probe technique detailed in Sec.~\ref{sec:main_result}. Specifically, we introduce a fictitious probe reaction, denoted $\hat{\rho}$, representing the direct interconversion $S \rightleftharpoons R$ (see the dashed red edge in Fig.~\ref{fig:chiral_breaking_fig3}a). Its stoichiometric vector is $\bS_{\hat{\rho}}=[1,-1,0,0]^T$ for the species $(R,S,A,C)$ respectively, thus the stoichiometric vector associated with internal species only, i.e., $R$ and $S$, is $\bS_{\hat{\rho}}^X=[1,-1]^T$. The inclusion of this probe into the network as an additional reaction 
leads to new EFMs involving the probe $\hat{\rho}$:
\begin{equation} 
\hat{\bm{e}}_1=[2,0,1,1]^T,  \hat{\bm{e}}_2 =[1,-1,0,1]^T, \hat{\bm{e}}_3 =[0,-2,-1,1]^T.
\end{equation}
These EFMs are represented in Fig.~\ref{fig:chiral_breaking_fig3}b-c. The affinity of the probe at steady state, $A_{\hat{\rho}}^{\rm ss} = \mu_S^{\rm ss} - \mu_R^{\rm ss}$, exactly quantifies chiral imbalance and 
is bounded by the dissipation along these EFMs, as dictated by Eq.~\eqref{eq:probe_affinity_bound}. Given that the standard chemical potentials of enantiomers are equal ($\mu_R^\circ = \mu_S^\circ$), this affinity directly translates to the concentration ratio: $\mu_R^{\rm ss} - \mu_S^{\rm ss} = RT\ln(r^{\rm ss}/s^{\rm ss})$. These specific EFMs yield bounds for $A_{\hat{\rho}}^{\rm ss}$ of $\pm \Delta\mu$ and $0$. Consequently, by using Eq.~\eqref{eq:eq_constant_bound}, we directly obtain rigorous thermodynamic bounds on the ratio $r^{\rm ss}/s^{\rm ss}$, thereby defining the thermodynamically accessible space for chiral imbalance: 
\begin{equation}\label{eq:SCSB_bound}
    -\Delta\mu/RT\leq\ln\frac{r^{\rm {ss}}}{s^{\rm ss}}\leq \Delta\mu/RT \;.
\end{equation}

The implications are illustrated in Fig.~\ref{fig:chiral_breaking_fig3}(d-f). An increase in the driving force $\Delta\mu$ expands the accessible thermodynamic space for concentrations of chiral species, depicted by the green shaded regions in the phase portraits (Fig.~\ref{fig:chiral_breaking_fig3}(d,e)) and the bifurcation diagram (Fig.~\ref{fig:chiral_breaking_fig3}(f)). The system undergoes a pitchfork bifurcation from an achiral state ($r^{\rm ss} = s^{\rm ss}$) to two stable chiral states ($r^{\rm ss} \neq s^{\rm ss}$), which are strictly confined by our thermodynamic bounds in Eq.~\eqref{eq:SCSB_bound}. As shown in Fig.~\ref{fig:chiral_breaking_fig3}(f), the extent of chiral symmetry breaking, quantified by $\ln(s^{\rm ss}/r^{\rm ss})$, emerges at a critical driving strength and enlarges within the thermodynamic limits defined by $\pm \Delta\mu/RT$ (red and blue boundary lines).

{Whether these limits are merely approached or actually saturated is set by kinetics alone. The symmetry-broken branch can be obtained analytically (see Appendix~\ref{app:chiral_solution}) by solving a quadratic equation that depends on the ratios $k^+_0/k^-_0$ and $k^+_1/k^-_1$, and on the
chemostatted concentrations of $A$ and $C$. The gap between chiral imbalance and thermodynamic limits then reads:
\begin{equation}
    \frac{\Delta\mu}{RT} - \ln\frac{r^{\rm ss}}{s^{\rm ss}} \xrightarrow{\textrm{large driving}} \ln\frac{\kappa}{\kappa-1} \;,
\end{equation} 
with $\kappa=k_1^+/k_0^-$, fixed solely by the dimerization-to-autocatalysis timescale separation. 
Making $R+S\rightleftharpoons C$ fast, i.e., $\kappa\to\infty$, while keeping $k_1^+/k_1^-$, and hence $\Delta\mu$, fixed, pushes the system towards a pseudo-equilibrium state with vanishing affinity on the dimerization step. Therefore, the stable chiral branches saturate the bound $\pm\Delta\mu/RT$, as shown in Fig.~\ref{fig:chiral_breaking_fig3}(f), in analogy with the saturation condition by pseudo-equilibrium pathways discussed in \cite{qureshi2026Thermodynamic} in the case of linear networks.}


This thermodynamic analysis reveals a key insight: the emergence of perfect homochirality (i.e., $s^{\rm ss} \approx 0$ or $r^{\rm ss} \approx 0$) necessitates an infinite driving force $\Delta\mu$. The framework, exemplified here with a minimal model, can be extended to more complex autocatalytic CRNs pertinent to origin of life investigations, crucially without any limitation on network topologies. Our results quantitatively connect the degree of non-equilibrium driving, manifested in cycle affinities, to the achievable extent of chiral selection. This approach thereby bridges non-equilibrium thermodynamics with symmetry breaking phenomena and offers a robust tool for assessing the thermodynamic costs and inherent constraints of order emergence in chemical evolution.

\subsection{Dissipative Self-Assembly\label{sec:self-assembly}}
\begin{figure*}[!tbh]
    \includegraphics[width=2\columnwidth]{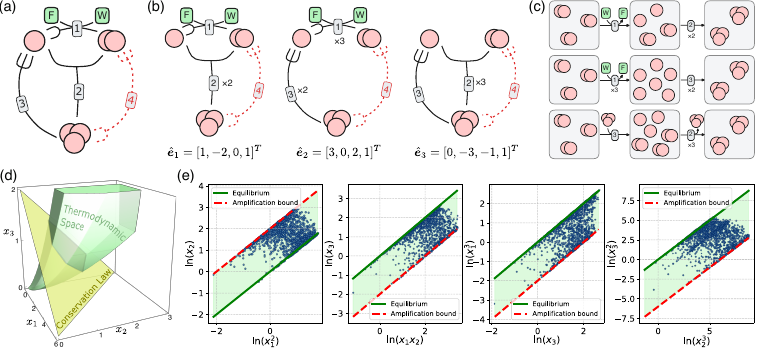}
    \caption{\textbf{Thermodynamic space for self-assembly.} (a) Reaction hypergraph of the self-assembly model showing three internal species (monomer, dimer, and trimer), two external species (F, W), and a chemical probe (dashed red edge). (b) Three Elementary flux modes (EFMs) including the chemical probe, with vector representation $\hat{\bm{e}}$. (c) Sequences of reactions through the original network corresponding to the three EFMs in (b), converting species between the two sides of the probe reaction. (d) Thermodynamic space (green) intersects with the conservation law plane (yellow), defining possible steady-state concentrations of internal species. (e) The thermodynamic space (green shaded area) computed from Eqs.~\eqref{eq:self_assembly_bound} and \eqref{eq:self_assembly_probe_bound}, where the green boundary represents equilibrium ratio, while the dashed red boundary establishes limits on non-equilibrium amplification. Dots show steady-state concentrations from simulations using random kinetic rates and total concentration at fixed thermodynamic driving $\Delta\mu\equiv \mu_F-\mu_W=2RT$ and standard chemical potential $\mu_{X_1}^\circ=\mu_{X_2}^\circ=\mu_{X_3}^\circ=0$.}
    \label{fig:self_assembly}
\end{figure*}

Self-assembly of complex macromolecules is fundamental to biological functions and advanced material design. At equilibrium, high-energy or kinetically stable structures, though often desired, may be poorly populated as the system tends to minimize its free energy. Driving the assembly process far from equilibrium can dramatically enhance the yield and specificity of these states \cite{rosenberger2021SelfAssembly,riess2020Designa}. Nevertheless, predicting the performance limits imposed by such non-equilibrium conditions remains a significant challenge. A robust framework is needed to quantitatively relate energy input to achievable abundance and complexity of given target assemblies. In this application, we employ our thermodynamic space formalism on a minimal self-assembly model. We demonstrate its capacity to define rigorous bounds on the amplification and attainable concentrations of complex structures.

The system, depicted in Fig.~\ref{fig:self_assembly}(a), involves three internal species: a monomer $X_1$, a dimer $X_2$, and a trimer $X_3$ with three reactions: a fuel-to-waste conversion driving the dimer formation, a spontaneous association of a dimer and a monomer to form a trimer, and the spontaneous disassembly from a trimer to three monomers. Here, the fuel $F$ and the waste $W$ are treated as chemostatted external species, keeping the system out of equilibrium. The detailed reaction scheme is illustrated in Eq.~\eqref{eq:self_assembly_reac}, together with the stoichiometric matrix, while the corresponding reaction hypergraph is represented in Fig.~\ref{fig:self_assembly}a with black hyperedges.
\begin{equation}\label{eq:self_assembly_reac}
    \begin{aligned}
      \rboxed{1}\  &F+ 2 X_1\xrightleftharpoons[k_1^-]{k_1^+} X_2 +W \;,\\
       \rboxed{2}\ &X_1 + X_2\xrightleftharpoons[k_2^-]{k_2^+} X_3 \;,\\
       \rboxed{3}\  &X_3\xrightleftharpoons[k_3^-]{k_3^+} 3X_1 \;.\\
    \end{aligned}
        \bS   
      =\left[
  \begin{array}{ccc}
    -2&-1&3 \\
    1&-1&0\\
    0&1&-1 \\
    \hline 
    -1 & 0 &0\\
    1 & 0 &0\\
  \end{array}
\right]
  \begin{array}{c}
    X_1 \\
    X_2\\
    X_3 \\
    \hline 
    F\\
    W\\
  \end{array}
\end{equation}
For this self-assembly CRN, there is one EFM, $\bm{e} = [1,1,1]^T$
in the reaction hypergraph and its cycle affinity, $A_{\bm{e}} = \mu_F-\mu_W = \Delta\mu>0$, stems from the fuel-to-waste driven dimer formation. We further assume that this chemical potential difference is strictly positive in the model. Applying our framework, the affinity of each reaction $\rho \in \{1,2,3\}$ at steady state is bounded according to Eq.~\eqref{eq:affinity_bound_1} as follows: 
\begin{equation}
   0\leq A_\rho^{\rm ss} \leq \Delta\mu, \text{ for $\rho = 1,2,3$}.
\end{equation}
Using Eq.~\eqref{eq:eq_constant_bound},  we can also bound the concentration ratios of internal species involved in these three reactions
\begin{equation}\label{eq:self_assembly_bound}
    \begin{aligned}
    e^{-\frac{\mu_{2}^{\circ}-2\mu_{1}^{\circ}}{RT}} \leq 
    &\frac{x_2^{\rm ss}}{(x_1^{\rm ss})^2}
    \leq  e^{-\frac{\mu_{2}^{\circ}-2\mu_{1}^{\circ}-\Delta\mu}{RT} }\\ 
 e^{-\frac{\mu_{3}^{\circ}-\mu_{2}^{\circ}-\mu_{1}^{\circ}+\Delta \mu}{RT} }    \leq 
 &\frac{x_3^{\rm ss}}{x_1^{\rm ss}x_2^{\rm ss}}
 \leq  e^{-\frac{\mu_{3}^{\circ}-\mu_{2}^{\circ}-\mu_{1}^{\circ}}{RT} } \\
  e^{-\frac{3\mu_{1}^{\circ}-\mu_{3}^{\circ}+\Delta\mu}{RT} }  \leq 
  &\frac{(x_1^{\rm ss})^3}{x_3^{\rm ss}}
  \leq e^{-\frac{3\mu_{1}^{\circ}-\mu_{3}^{\circ}}{RT} }  \\
    \end{aligned}
\end{equation}
where we use $\mu_1^{\circ}$, $\mu_2^{\circ}$, and $\mu_3^{\circ}$ to denote the standard chemical potential of internal species $X_1$, $X_2$ and $X_3$, respectively. The constraints on concentrations define the thermodynamic space illustrated in Fig.~\ref{fig:self_assembly}d. 

Additionally, to determine the accessible region of internal species concentrations, we also have to consider the existing conservation laws (see Fig.~\ref{fig:self_assembly}d). This model has one elementary conservation law (ECL) derived from the right nullspace of $(\bS^X)^T$, given by $\bm{r} = [1,2,3]^T$. This implies the conservation of the total number of monomeric units, $x_1+2x_2+3x_3 = \text{constant}$. This conservation law permits an additional effective reaction. 
\begin{equation}
 \rboxed{4}\ 3X_2 \xrightleftharpoons[k_4^-]{k_4^+} 2 X_3
\end{equation}
which is not explicitly in the original CRN but belongs to the reaction space. Therefore, we can introduce this reaction through a chemical probe (Fig.~\ref{fig:self_assembly}a-b). 
This corresponds to the addition of a new stoichiometric column  $\hat{S}_{\hat{\rho}} = [0,-3,2,0,0]^T$ which is, by definition, orthogonal to the conservation law, i.e., $\bm{r}^T\hat{S}_{\hat{\rho}}^X = 0$. The extended stoichiometric matrix $\tilde{\bS}= [\bS,\hat{\bS}_{\hat{\rho}}]$ exhibits three EFMs 
that enclose the extra reaction given by the chemical probe:
\begin{equation*}
     \hat{\bm{e}}_1 =[1, -2, 0, 1]^T,\ 
     \hat{\bm{e}}_2 =[3, 0, 2, 1]^T,\ 
     \hat{\bm{e}}_3 =[0, -3, -1, 1]^T
\end{equation*}
illustrated in Fig.~\ref{fig:self_assembly}b.
From this, we can find the bounds on species concentrations on the two sides of the chemical probe, i.e., reaction $4$, by using Eq.~\eqref{eq:eq_constant_bound}. We first find the pathways that convert the dimer into the trimer from this set of EFMs, and then evaluate their corresponding maximum and minimum (pseudo-)equilibrium constants:
\begin{equation}\label{eq:self_assembly_probe_bound}
    e^{-\frac{2\mu_{3}^{\circ}-3\mu_{2}^{\circ}+3\Delta\mu}{RT}}\leq \frac{(x_3^{\rm ss})^2}{(x_2^{\rm ss})^3}\leq   e^{-\frac{2\mu_{3}^{\circ}-3\mu_{2}^{\circ}}{RT}} \;.
\end{equation}
Analogously, we can bound the effective affinity of this additional reaction by means of Eq.~\eqref{eq:probe_affinity_bound}:
\begin{equation}
    0\leq  A_{\hat{\rho}}^{\rm ss}\leq  3\Delta\mu
\end{equation}

Figure~\ref{fig:self_assembly}e depicts this thermodynamic space explicitly for various combinations of species concentration, as indicated by Eq.~\eqref{eq:self_assembly_bound} and \eqref{eq:self_assembly_probe_bound}. In this example, one side of the boundary represents the equilibrium value determined through standard chemical potential differences, while the other side dictates the limits on non-equilibrium amplification for a given driving $\Delta\mu$. The scattered points indicate steady-state concentrations from numerical simulations at a fixed $\Delta\mu = 2RT$, with randomized kinetic parameters and total concentrations. All points fall within the predicted bounds, validating our theoretical predictions. Although this example uses a minimal model, it clearly demonstrates a key principle: the interplay between non-equilibrium driving (via $\Delta\mu$) and network topology (reflected in EFMs and conservation laws) establishes a well-defined accessible thermodynamic space for the stationary concentrations in a self-assembly system. This framework further provides rigorous bounds on the extent to which the abundance of complex structures (e.g., trimers in this example) can be amplified beyond their equilibrium values.

Our approach can be directly applied to more complex self-assembly models involving multiple steps and intermediate states, offering a universal bound on the capabilities of dissipative self-assembly from a purely topological and thermodynamic perspective. This offers a powerful tool for understanding structural principles underpinning natural self-assembling systems and guiding the design of synthetic ones, with a particular focus on predicting how energy input can be harnessed to achieve specific structural or functional outcomes.

\begin{figure*}[t]
    \includegraphics[width=2\columnwidth]{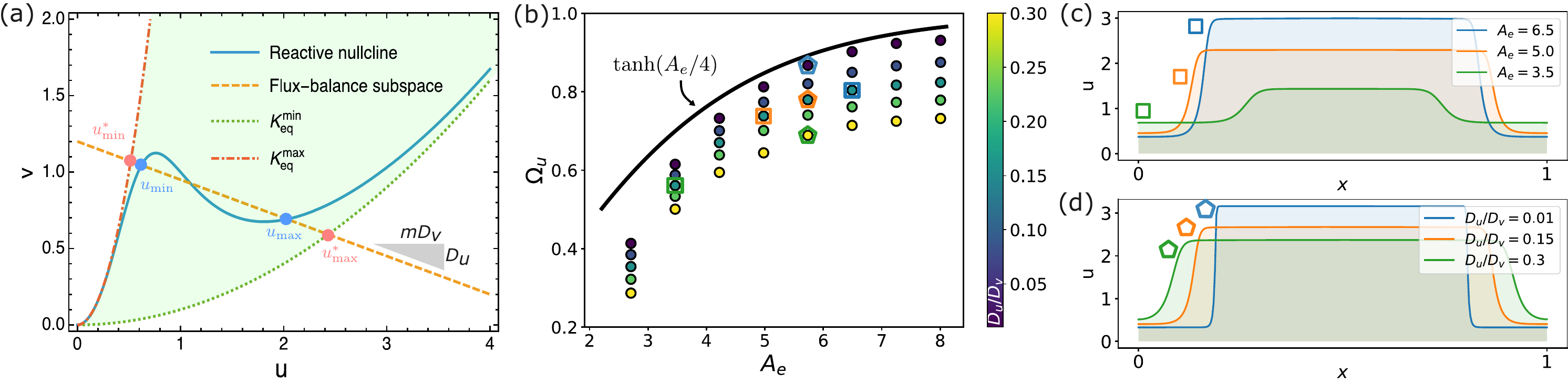}
    \caption{\textbf{Thermodynamic constraints on reaction-diffusion patterns.} (a) The phase-space of concentrations shows that the reactive nullcline is embedded in the thermodynamic space, whose boundaries are determined by $K_{\rm eq}^{\rm min}$ and $K_{\rm eq}^{\rm max}$ (see Eq.~\eqref{eq:nullcline_bound}). The intersection between thermodynamic space and the flux-balance subspace gives the range of concentration of stationary patterns. {\color{black}This illustrative panel uses $m=2$ and $n=4$, with the reactive nullcline obtained by setting $f(u,v)=0$ in Eq.~\eqref{eq:reaction_kinetics}.} (b) Pattern contrast $\Omega_u$ versus cycle affinity $A_e=\mu_F-\mu_W=\Delta\mu$. The theoretical maximum contrast is the black curve, while the contrast from numerical simulations (circles) has been obtained for various diffusion coefficient ratios $D_v/D_u$. The theoretical line is approached for $D_u/D_v \to 0$. (c) Stationary concentration profiles, {obtained for three cycle
    affinities at a fixed diffusion coefficient ratio
    $D_u/D_v$ corresponding to square points in panel (b), demonstrate} how pattern amplitude increases with driving force, always remaining bounded by thermodynamic constraints. (d) {Stationary concentration profiles, obtained for three
    diffusion coefficient ratios at a fixed
    cycle affinity $A_e$ (pentagons in panel (b)).} In the numerical simulation {\color{black}of panels (b--d)}, we set $D_v=10$, $k_+=0.9$, $k_-=0.3$, and $w_-[F]=1$, while varying $w_+[W] \in [0.001, 0.2]$ and $D_u \in [0.1, 3]$.}
    \label{fig:Rd pattern}
\end{figure*}


\subsection{Reaction-diffusion pattern}
Reaction-diffusion patterns are paradigmatic examples of dissipative self-organization, and quantifying their thermodynamic cost is a central challenge in non-equilibrium thermodynamics \cite{zhang2023Free,falasco2018Information,nagayama2025Geometric}. In this final application, we extend our thermodynamic space framework from well-mixed systems to these spatially extended phenomena. While our previous work provided bounds for pattern-forming catalytic networks \cite{liang2024Thermodynamic}, we now generalize this analysis to arbitrary reaction topologies. This allows us to establish universal thermodynamic constraints on pattern contrast, revealing how the network stoichiometric structure fundamentally limits macroscopic spatial organization.

We investigate a reaction-diffusion system that exhibits non-equilibrium pattern formation through a minimal two-species reaction network. The system features two competing pathways of conversion between $U$ and $V$:
\begin{equation}\label{eq}
\begin{aligned}
mU &\xrightleftharpoons[k_-]{k_+} V \;,\\
W+(n+m)U&\xrightleftharpoons[w_-]{w_+} nU+V+F \;.
\end{aligned}
\end{equation}
The first reaction represents a spontaneous $m$-th order conversion between internal species. The second describes a fuel-driven catalytic process where $n$ molecules of $U$ act as catalysts, facilitating the conversion of $m$ additional molecules of $U$ into $V$ and vice-versa. This catalytic pathway maintains the system out of equilibrium being coupled to the interconversion between chemostatted fuel species $F$ and waste species $W$.

The spatiotemporal dynamics is governed by coupled reaction-diffusion equations:
\begin{equation}\label{eq:rd_dynamics}
\begin{aligned}
\partial_t u &= D_u\nabla^2 u + mf(u,v),\\
\partial_t v &= D_v\nabla^2 v - f(u,v)
\end{aligned}
\end{equation}
where the reaction kinetics are given by:
\begin{equation}\label{eq:reaction_kinetics}
f(u,v) = -k_+u^m + k_-v - w_+'u^{n+m} +w_-'u^nv
\end{equation}
Here, $w_+' = w_+ [W]$ and $w_-' = w_-[F]$ incorporate the concentrations of chemostatted external species. The cycle affinity $A_e = \Delta\mu=\mu_F - \mu_W$ quantifies the non-equilibrium driving force maintaining the system away from detailed balance.

We focus on a mass-conserving reaction-diffusion system without boundary fluxes.
{By taking the appropriate stoichiometric linear combination of the steady-state reaction-diffusion equations in Eq.~\eqref{eq:rd_dynamics}, the reaction terms cancel and one obtains
\begin{equation}
\nabla^2 \eta^{\rm ss}=0 \;,
\qquad
\eta^{\rm ss}=u^{\rm ss}+m\frac{D_v}{D_u}v^{\rm ss} \;.
\end{equation}
Since $\eta^{\rm ss}$ is a harmonic function satisfying homogeneous Neumann boundary conditions on a connected domain, it must be constant. Thus,} the system obeys the following conservation law that constrains the spatial distribution:
\begin{equation}\label{eq:FB_subspace}
D_u u^{\rm ss} + mD_v v^{\rm ss} = \text{const},
\end{equation}
which defines a flux-balance subspace in the concentration space. {For one-dimensional systems, Eq.~\eqref{eq:FB_subspace} can be obtained by direct integration.} As illustrated in Fig.~\ref{fig:Rd pattern}(a), and following the phase-space geometry approach \cite{brauns2020PhaseSpace}, the intersection of this subspace with the reactive nullcline $f(u,v)=0$ determines the extremal concentrations explored by the stationary pattern. Our framework reveals that the reactive nullcline, {i.e., the set of concentrations for which the reaction term associated with the CRN in Eq.~\eqref{eq} vanishes,} is fundamentally constrained by the following thermodynamic bounds:
\begin{equation}\label{eq:nullcline_bound}
    \begin{aligned}
    \frac{v^{\rm ss}}{(u^{\rm ss})^m}\leq &K_{eq}^{\rm max}=\frac{k_+}{k_-}=e^{-\frac{\mu_v^\circ-m\mu_u^\circ}{RT}}\\
    \frac{v^{\rm ss}}{(u^{\rm ss})^m}\geq &K_{eq}^{\rm min}=\frac{w_+[W]}{w_-[F]}=e^{-\frac{\mu_v^\circ-m\mu_u^\circ+\Delta\mu}{RT}}\\
\end{aligned}
\end{equation}
By identifying the intersections of the flux-balance subspace, Eq.~\eqref{eq:FB_subspace} with the boundaries of the thermodynamic space, Eq.~\eqref{eq:nullcline_bound}, we can find bounds on the
accessible stationary concentrations in a stationary pattern of the
two internal species $U$ and $V$. In Fig.~\ref{fig:Rd pattern}(a), they are indicated as $u^*_{\rm min}$ and $u^*_{\rm max}$ for the species $U$. However, these bounds still depend on the diffusion coefficients, since the
slope of the flux-balance subspace is determined by $D_u/(mD_v)$. To obtain diffusion-independent bounds, we further maximize the accessible range of $U$ over all possible slopes of the flux-balance subspace. Since this slope is non-positive and approaches zero in the limit
$D_u/D_v\to0$, the widest interval is obtained in this limit. The flux-balance subspace {then becomes} a horizontal line and the bound on the range of the stationary concentrations of $U$ takes the maximum value. {In particular, from Eq.~\eqref{eq:nullcline_bound}, we obtain:
\begin{equation}
    \frac{u_{\rm max}^{\rm ss}}{u_{\rm min}^{\rm ss}}\leq \frac{u_{\rm max}^*}{u_{\rm min}^*}\Bigg|_{(D_u/D_v) \to 0}=e^{\Delta\mu/mRT} \;.
\end{equation}}
Therefore, in this limit, the pattern contrast satisfies:
\begin{equation}\label{eq:pattern_contrast_bound}
    \Omega_u = \frac{u_{\rm max}^{\rm ss}-u^{\rm ss}_{\rm min}
    }{u^{\rm ss}_{\rm max}+u^{\rm ss}_{\rm min}}\leq \tanh\left[\frac{\Delta\mu}{2mRT}\right]
\end{equation}

The numerical verification of the bound is shown in Fig.~\ref{fig:Rd pattern}(b-d), where we take $m=2$ and $n=4$. The deviation from the theoretical bound is related to the diffusion coefficient ratio of the two species --- when the diffusion of species $v$ is much faster than that of $u$, almost all non-equilibrium driving responsible for the pattern contrast is allocated on the species $U$ and the bound in Eq.~\eqref{eq:pattern_contrast_bound} gets saturated in the limit $D_u/D_v \to 0$.

\section{Applications to experimental data and thermodynamic inference\label{sec:data}}

In this section, we present two {applications to experimental data}: in the first one, we test the validity of our bounds in a concrete CRN describing protein conformational changes; in the second example, we show that our results can be used to extract useful thermodynamic information directly from experimental data.
\begin{figure*}[!tbh]
    \centering
    \includegraphics[width=1\linewidth]{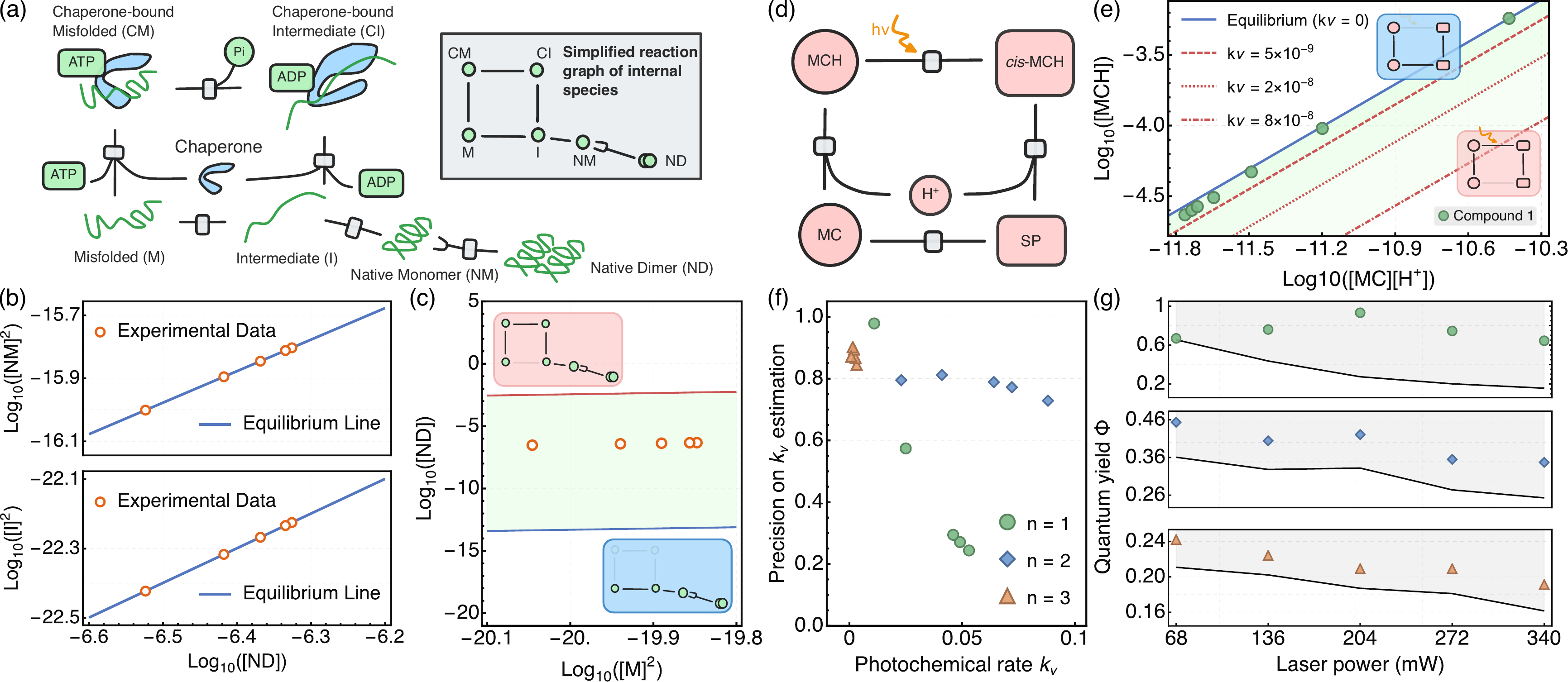}
    \caption{\textbf{Data-driven applications of thermodynamic space.} (a) CRN describing the action of a chaperone (GroEL) on MDH proteins (malate dehydrogenase). Native MDH dimers can misfold through an intermediate state, while chaperones use ATP to stabilize the native folded configurations. The gray box contains a simplified version of the reaction graph. (b) The concentration of native dimer ([ND]) against that of native monomer ([NM]) and intermediates ([I]). Experimental data (dots) stay on the equilibrium line even in the presence of ATP ($\Delta \mu_{\rm ATP} = 7.7 \,\rm kcal/mol$). (c) The concentration of misfolded proteins ([M]) is displaced from equilibrium with respect to [ND] and lies within the thermodynamic space (green area). Blue and red lines indicate lower and upper bounds with respective pathways (colored boxes). Parameters have been obtained from \cite{goloubinoff2018Chaperones}. (d) CRN describing a light-driven photoacid. Upon irradiation (yellow arrow), the transition from MCH to its cis isomer is accelerated by a rate $k_\nu$, producing a net pH change in the solution. (e) MCH concentration ([MCH]) against [MC][$\rm H^+$] for the compound 1, i.e., MCH with a chemical bridge of length $n=1$. Data from \cite{berton2020thermodynamics} are not compatible with the equilibrium line, but they lie within the thermodynamic space (green area) plotted for increasing nonequilibrium driving $k_\nu$. Colored boxes represent the pathways leading to upper and lower bounds. (f) Precision of $k_\nu$ estimation obtained as $1-\epsilon_\nu$, with $\epsilon_\nu = |1 - k^{\rm exp}_{\nu}/k^{\rm min}_{\nu}|$, the relative error between experimental and inferred minimal values of $k_\nu$ compatible with observed concentrations. Different values of $n$ indicate different compounds. (g) From $k_\nu^{\rm min}$, we show the lower bound (black line) on the power-dependent quantum yield $\Phi$ for each compound. Parameters have been obtained from \cite{berton2020thermodynamics}.}
    \label{fig:data}
\end{figure*}

\subsection{Stabilization of native proteins by chaperones}

Proteins operate their functions when found in their native conformations \cite{dobson2003protein}. However, as a consequence of shocks or environmental stress, they can undergo denaturation, populating deep local minima of their free energy landscape from which it is difficult to spontaneously escape, and eventually form stable aggregates, such as beta-amyloid and alpha-synuclein fibrils \cite{dobson2003protein,amartumur2024neuropathogenesis,goloubinoff2018Chaperones}. Since misfolded proteins and aggregated conformations have been identified as primary causes for several neurodegenerative diseases, they recently attracted a lot of attention \cite{hipp2014proteostasis,goloubinoff2016mechanisms}. In this scenario, chaperone proteins play a leading role, being able to refold misfolded proteins by using energy from ATP, leading to an increased nonequilibrium stabilization of the native state \cite{goloubinoff2018Chaperones}.

In Fig.~\ref{fig:data}a, we present the CRN describing the action of GroEL (chaperone) on MDH proteins (malate dehydrogenase), ignoring the presence of aggregates due to the presence of chaperones \cite{goloubinoff2018Chaperones}. By extracting experimental data and parameter values on this system from \cite{goloubinoff2018Chaperones}, we computed the thermodynamic bound for different pairs of concentrations, $[\rm ND]$ (native dimers) versus $[\rm NM]^2$ (native monomers), $[\rm ND]$ versus $[\rm I]^2$ (intermediates), and $[\rm ND]$ versus $[\rm M]^2$ (misfolded proteins). In Figs.~\ref{fig:data}b-c, we show these ratios for different total protein concentrations but the same amount of ATP, i.e., we change the constraints imposed by the conservation laws while keeping the nonequilibrium driving fixed. 
{\color{black}The total protein concentrations are extracted from Ref.~\cite{goloubinoff2018Chaperones} and account, at a coarse-grained level, for the loss of soluble protein due to aggregation before chaperones are introduced. We then add a fixed amount of chaperones.} We show that, while $r_{\rm ND,NM}$ and $r_{\rm ND, I}$ cannot be displaced from their equilibrium value (Fig.~\ref{fig:data}b), the ratio of native dimers over misfolded proteins is much higher than at equilibrium (Fig.~\ref{fig:data}c) and all data fall into the thermodynamic space analytically predicted. This result indicates that the only role of ATP hydrolysis is to effectively increase the free energy of misfolded proteins through the chaperone machinery, by an amount constrained by the energetic budget. This observation is in complete agreement with the scenario outlined in \cite{goloubinoff2018Chaperones}.

\subsection{Light-driven photoacids}

The second example concerns the dynamics of the protonated merocyanine in water. This is a photoacid that undergoes a sequence of structural and acid-base transformations, as illustrated in Fig.~\ref{fig:data}d. Starting from its protonated form (${\rm MCH}$), the system can deprotonate to yield the merocyanine (${\rm MC}$) {\color{black}with a reaction rate constant $k_1$} and release a ${\rm H}^+$ in solution. Through a ring-closure reaction, ${\rm MC}$ can convert into the spiropyran (${\rm SP}$) {\color{black}with a reaction rate constant $k_2$}, which in turn can be protonated to form the cis isomer of MCH ($\textit{cis-}{\rm MCH}$) {\color{black}with a reaction rate constant $k_3$}. This species eventually converts back into MCH {\color{black}with a reaction rate constant $k_4$}, closing the reaction cycle. {\color{black}The reaction rate constants associated with the reverse cycle are $k_{-i}$, with $i = 1, \dots, 4$.} Under light irradiation, the reaction from ${\rm MCH}$ to $\textit{cis-}{\rm MCH}$ is accelerated, {while being extremely slow in the absence of light. Following \cite{berton2020thermodynamics}, and for the sake of simplicity, we assume that light irradiation increases the reaction rate from ${\rm MCH}$ to its cis isomer by} a photochemical rate $k_\nu$, driving the system into a photo-stationary state that leads to a measurable change of pH in the solution. Owing to this property, these chemical species can be employed as tunable pH buffers \cite{berton2021LightSwitchable}. This cyclic set of operations has been first characterized both experimentally and theoretically in \cite{berton2020thermodynamics} for a family of ${\rm MCH}$ featuring chemical bridges of different lengths, $n=1,2,3$, and visible-light sources of different powers. This characterization provides a reliable ground-truth against which we can validate the predictions obtained from our bounds. We remark that, to make such a comparison, as said above, we are following the model introduced in the literature, thereby not considering photon-free and photon-assisted pathways as separate entities. 
{\color{black}As a consequence, the obtained results should be interpreted as thermodynamic consistency bounds within this effective coarse-grained model.} Further work, starting from experimental data, might be dedicated to this extended analysis {\color{black}and to an explicit thermodynamic description of light-driven photoacids.}

From the experiments in \cite{berton2020thermodynamics}, in Fig.~\ref{fig:data}e, we report $[{\rm MCH}]$ versus $[{\rm MC}][{\rm H}^+]$ in different photo-stationary conditions, each one associated with a different laser power. We show that these concentrations are bounded by the equilibrium constants associated with the two reaction pathways converting ${\rm MC}+{\rm H}^+$ into ${\rm MCH}$, according to Eq.~\eqref{eq:eq_constant_bound}. The equilibrium line is not compatible with the experimental data, while the accessible space enlarges as we increase the nonequilibrium driving $k_\nu$.

\paragraph*{Inference of nonequilibrium driving.} Imagine to perform a different, and simpler, experiment. We observe a photoacid in solution, whether $n = 1$, $2$, or $3$, exposed to a visible-light source whose power $P$ is not known. From UV-vis absorbances at selected wavelengths, it is easy to extrapolate ratios of stationary concentrations of all species in solution, $[{\rm MCH}]$, $[{\rm MC}]$, $[{\rm SP}]$, and $[\textit{cis-}{\rm MCH}]$. From Eq.~\eqref{eq:eq_constant_bound}, focusing on one single ratio for simplicity, we have:
\begin{equation}
    K^{\rm min}_{\rm MC \to MCH} [{\rm H}^+] \leq \frac{[{\rm MCH}]^{\rm obs}}{[{\rm MC}]^{\rm obs}} \leq K^{\rm max}_{\rm MC \to MCH} [{\rm H}^+]
    \label{eq:bound_H+}
\end{equation}
where the superscript $\cdot^{\rm obs}$ indicate observed quantities.

In this setting, we are not able to measure $[\rm H^+]$ through the pH to directly verify the bounds. However, pH is a function of $P$, ${\rm pH}(P)$. From previous calibration experiments performed independently (e.g., see \cite{berton2020thermodynamics}), we infer the function ${\rm pH}(P)$ -- and hence $[{\rm H}^+](P)$ -- through a fitting procedure of available data. Inserting $[{\rm H}^+](P)$ into Eq.~\eqref{eq:bound_H+}, {\color{black}and combining it with the equilibrium estimate 
$K_a=k_1/k_{-1}=10^{-{\rm p}K_a}$ for the 
${\rm MCH}\rightleftharpoons{\rm MC}+{\rm H}^+$ reaction,} we obtain the minimum value of power, $P^{\rm min}$, such that upper and lower bounds hold. In particular, we found that the upper bound, {\color{black}$K^{{\rm max}}_{{\rm MC \to MCH}} [{\rm{H}}^+] = 10^{{\rm p}K_a} [{\rm{H}}^+]$,} provides the more informative constraint, allowing us to determine the minimum nonequilibrium driving compatible with the observations, obtained at an unknown power. Experiments in conditions of controlled power also provide an estimation of average quantum yield, $\Phi$, for each compound. Since $\Phi = \alpha k_\nu$ (with $\alpha$ quantifying photon energy per molecule \cite{berton2020thermodynamics}), we map this minimum power into a minimum value of $k_\nu$, $k_\nu^{\rm min}$, compatible with our observations.

To test the proposed inference of $k^{\rm min}$, we employ it for the values of $[\rm MCH]^{\rm obs}$ and $[\rm MC]^{\rm obs}$ in \cite{berton2020thermodynamics}. In this case, we know the correct value of all parameters, in particular the ground-truth for $k_\nu$, $k_\nu^{\rm exp}$, and thus we are able to quantify the robustness of our approach. In Fig.~\ref{fig:data}f, we plot the relative error between $k_{\nu}^{\rm exp}$ and $k_\nu^{\rm min}$ for the three compounds at different laser powers. We notice a remarkably good performance of the inference method, with up to $75-85\%$ of nonequilibrium driving being captured by the thermodynamic bound in Eq.~\eqref{eq:bound_H+}. Experimentally, slightly different quantum yields have been found at different powers. Clearly, the estimation of $k^{\rm min}$ leads to a lower bound for the power-dependent $\Phi$ for each compound (black solid lines in Fig.~\ref{fig:data}g). These results provide indication that our thermodynamic bounds, in principle, can be employed to extrapolate information about nonequilibrium drivings in CRNs.

\section{Discussion}
In this work, we have derived fundamental thermodynamic constraints on generic CRNs at stationarity, introducing the novel concept of \textit{thermodynamic space} to describe the accessible range of species concentrations compatible with the system's energy budget. Similar bounds can also be obtained for reaction affinities and employed to estimate the limits of effective affinities between any two species in the CRN. Our approach is based on CRN hypergraph geometry and utilizes the novel \textit{chemical probe} technique to study species that are not directly linked by an existing reaction. Therefore, it significantly extends and generalizes previous findings on linear and catalytic networks to the challenging case of arbitrary network topologies. This generalization is crucial for applications to complex multimolecular CRNs, such as multi-stationary biochemical networks and self-assembly processes, which are widespread in living systems, but their investigations have fallen short of comprehensive thermodynamic analyses. We believe that the results we introduced will be crucial to characterize CRNs starting solely from global thermodynamic properties, without the need to rely on detailed knowledge of the entire kinetics.

We demonstrate our framework's applicability through four examples, the Schl\"ogl model for bistability, a thermodynamically consistent model for chiral symmetry breaking, a simple self-assembly amplification, and the onset of reaction-diffusion patterns. These case studies illustrate how our bounds elucidate thermodynamic prerequisites for the emergence of various complex behaviors in biological and biochemical systems. Our approach bridges concepts from non-equilibrium thermodynamics, network theory, and systems biology, offering a powerful interdisciplinary tool for analyzing living systems. We have also developed an open-source Python package, TACOS, to determine the Thermodynamic Space of arbitrary small- to medium-size CRNs (see Appendix \ref{sec:TACOS}) with the aim of facilitating the broad application of the presented results.

The generality of our framework opens avenues for diverse applications beyond those explored here. For instance, it could be extended to bound responses to kinetic parameter perturbations in multi-molecular CRNs, building upon previous results for linear systems \cite{owen2020Universal}. Our approach also has the potential to generalize studies on the information propagation in molecular templating networks without making pseudo-unimolecular constraints \cite{qureshi2026Thermodynamic,sartori2015Thermodynamics}. Furthermore, the idea of building a thermodynamic space has been previously discussed in the context of metabolic networks as a constrained optimization problem \cite{chakrabarti2013Kinetic}. Our framework provides the exact form of this space, thereby converting the nonlinear, high-dimensional problem of optimizing over kinetic rate constants into a structural thermodynamic identification of optimal reaction pathways. Moreover, the derived bounds do more than simply delimit the stationary solutions: they reveal the chemical pathways through which optimality is achieved. These results may therefore help identify foundational building blocks for characterizing the operating range of metabolic activity in any CRN, based on thermodynamic considerations involving free energies and externally chemostatted species. Future works could also explore applications to other non-linear phenomena in biochemical systems, such as biochemical oscillations. An exciting direction would be to investigate thermodynamic bounds on various attractor types (e.g., limit cycles, strange attractors), potentially allowing to extend the concept of thermodynamic space beyond stationary states. This could provide insights into the thermodynamic requirements for complex dynamical behaviors in living systems, such as circadian rhythms and calcium oscillations \cite{cao2015Freeenergy,voorsluijs2024Calcium}.

In conclusion, this work provides a general framework for analyzing non-equilibrium phenomena in living and synthetic systems, contributing to a deeper understanding of the principles governing complex chemical and biological processes. By establishing fundamental thermodynamic constraints on the operation of CRNs, our approach offers new perspectives on the design principles of living systems and can potentially guide the development of artificial molecular systems with life-like properties, such as the emergence of informational sequences through complex self-assembly processes. 
Ultimately, this work represents a significant step towards a more comprehensive thermodynamic theory of complex, far-from-equilibrium, biological and biochemical systems supporting emergent functional behaviors.







\begin{acknowledgements}
S.L. thanks Massimo Bilancioni for stimulating discussions on elementary flux modes and acknowledges helpful discussion with Prof.~Vassily Hatzimanikatis. D.M.B. thanks David Lacoste for critical reading of the manuscript and enlightening discussions on the results. S.L. and P.D.L.R. thank the Swiss National Science Foundation for support under grant 200020\_178763 and CRSII5\_193740. Part of this work was finished by S.L. and D.M.B. during the Frontiers in Non-equilibrium Physics 2024 Workshop at Yukawa Institute for Theoretical Physics, Kyoto University. D.M.B. is funded by the program STARS@UNIPD with the project ``ActiveInfo''.
\end{acknowledgements}

\appendix

\dmb{\section{Exact solution of the chiral symmetry-breaking model and its limiting behavior}\label{app:chiral_solution}}

{
The thermodynamically consistent Frank model introduced in the main text (see Eq.~\eqref{eq:frank_reaction}) is exactly solvable at steady state. We collect here its closed-form fixed points, the resulting chiral imbalance, and the limiting behavior that underlies the saturation of the bound shown in Fig.~\ref{fig:chiral_breaking_fig3}(f). With the achiral precursor $A$ and the dimer $C$ chemostatted at $a\equiv[A]$ and $c\equiv[C]$, respectively, mass-action kinetics for the enantiomer concentrations $r\equiv[R]$ and $s\equiv[S]$ read
\begin{equation}\label{eq:frank_ode}
\begin{aligned}
\dot r &= k_0^+ a\,r - k_0^- r^2 - k_1^+ r s + k_1^- c \;,\\
\dot s &= k_0^+ a\,s - k_0^- s^2 - k_1^+ r s + k_1^- c \;.
\end{aligned}
\end{equation}
The exchange $R\leftrightarrow S$ leaves Eq.~\eqref{eq:frank_ode} invariant; this $\mathbb{Z}_2$ symmetry is what makes the model solvable.

\paragraph*{Fixed points analysis.}
Taking the difference between the two stationarity conditions, the symmetric coupling $-k_1^+ rs+k_1^- c$ cancels out and the concentration difference factorizes:
\begin{equation}\label{eq:frank_subtract}
0=\dot r-\dot s=(r-s)\big[k_0^+ a-k_0^-(r+s)\big].
\end{equation}
A steady state is therefore either \emph{achiral} ($r=s$) or \emph{chiral} with $r+s=k_0^+ a/k_0^-$. On the chiral branch, to close the system, we sum the two stationary conditions and, using $r^2+s^2=(r+s)^2-2rs$, we arrive at a linear equation for $rs$. Therefore, the enantiomer concentrations are the two roots of a quadratic equation, i.e.,
\begin{equation}\label{eq:frank_roots}
\begin{gathered}
r^{\rm ss},\,s^{\rm ss}=\tfrac12\!\left(P\pm\sqrt{P^2-4Q}\right),\\
P=\frac{k_0^+ a}{k_0^-},\qquad
Q=\frac{k_1^- c}{k_1^+-k_0^-}.
\end{gathered}
\end{equation}
In contrast, the achiral solution, obtained by setting $r=s$ in Eq.~\eqref{eq:frank_ode}, is
\begin{equation}\label{eq:frank_achiral}
r^{\rm ss}=s^{\rm ss}=\frac{k_0^+ a+\sqrt{(k_0^+ a)^2+4c\,k_1^-(k_0^-+k_1^+)}}{2(k_0^-+k_1^+)},
\end{equation}
and always exists. {However, the} chiral branch consists of two positive real roots if and only if $k_1^+>k_0^-$ (so that $Q>0$) and $P^2\ge 4Q$.

\paragraph*{Chiral imbalance and pitchfork bifurcation.}
{By introducing the parameter} $\beta\equiv 4Q/P^2$, Eq.~\eqref{eq:frank_roots} yields the exact order parameter
\begin{eqnarray}\label{eq:frank_lnratio}
\ln\frac{r^{\rm ss}}{s^{\rm ss}}&=&2\,\mathrm{artanh}\sqrt{1-\beta} \\
\beta&=&\frac{4k_1^+}{k_1^+-k_0^-} e^{-\Delta\mu/RT} \nonumber \;,
\end{eqnarray}
where local detailed balance fixes $\Delta\mu/RT=2\ln(k_0^+/k_0^-)+\ln(k_1^+/k_1^-)+\ln(a^2/c)$.
The chiral solution exists for $\beta<1$ and bifurcates continuously off the achiral branch at $\beta=1$, i.e., a supercritical pitchfork bifurcation is located at
\begin{equation}\label{eq:frank_onset}
\frac{\Delta\mu_{\rm crit}}{RT}=\ln\frac{4k_1^+}{k_1^+-k_0^-} \;.
\end{equation}
Linearizing Eq.~\eqref{eq:frank_ode} around $r=s$, {we find that the achiral state is stable for $\beta>1$ and a saddle for $\beta<1$, where the two chiral branches are the stable attractors. {In particular,} $\Delta\mu_{\rm crit}/RT\to\ln 4$ as $\kappa\equiv k_1^+/k_0^-\to\infty$.}

\paragraph*{Saturation of the bound.}
The probe bound in Eq.~\eqref{eq:SCSB_bound} confines the imbalance to $|\ln(r^{\rm ss}/s^{\rm ss})|\le\Delta\mu/RT$. Together with Eq.~\eqref{eq:frank_lnratio}, the distance to the bound reads
\begin{equation}\label{eq:frank_gap}
\begin{aligned}
\frac{\Delta\mu}{RT}-\ln\frac{r^{\rm ss}}{s^{\rm ss}}
&=\ln\frac{\kappa}{\kappa-1}
+\ln\!\left[\frac{4}{\beta}\,\frac{1-\sqrt{1-\beta}}{1+\sqrt{1-\beta}}\right]\\
&=\ln\frac{\kappa}{\kappa-1}+\frac{\beta}{2}+O(\beta^2) \;.
\end{aligned}
\end{equation}
Two consequences follow. First, at fixed kinetics, this gap decreases monotonically with the driving and tends to the nonzero constant $\ln[\kappa/(\kappa-1)]$ as $\Delta\mu\to\infty$ ($\beta\to0$): the chiral imbalance approaches the bound but never reaches it, so complete homochirality ($|\ln r^{\rm ss}/s^{\rm ss}|\to\infty$) requires infinite driving. Second, this asymptotic gap is exactly the steady-state affinity of the dimerization $R+S\rightleftharpoons C$,
\begin{equation}\label{eq:frank_A3}
\begin{aligned}
\ln\frac{\kappa}{\kappa-1}=\frac{A_3^{\rm ss}}{RT}
&=\ln\frac{k_1^+ r^{\rm ss}s^{\rm ss}}{k_1^- c} \;.
\end{aligned}
\end{equation}
The separation $\kappa$ is thus a purely kinetic knob: scaling $k_1^\pm$ together at fixed ratio $k_1^+/k_1^-$ increases $\kappa$ while leaving $\Delta\mu$ --- and hence the bound itself --- unchanged. As $\kappa\to\infty$, the dimerization reaches equilibrium ($r^{\rm ss}s^{\rm ss}\to (k_1^-/k_1^+)\,c$ ): the pathway determining the bound carries vanishing affinity, and the bound is therefore saturated, $\ln[\kappa/(\kappa-1)]\to0$. This realizes, for a genuine stoichiometric-hypergraph CRN, the pseudo-equilibrium saturation condition discussed for linear networks in Ref.~\cite{qureshi2026Thermodynamic}. This behaviour is captured by Fig.~\ref{fig:chiral_breaking_fig3}(f).}

\section{Enumeration of Elementary Flux Modes and its scaling}\label{sec:EFM_enumeration}

In the main text, cycles (flux modes) are denoted by vectors $\bm{c}$ satisfying $\bS^X\bm{c}=\bm{0}$, while elementary flux modes (EFMs) are defined as minimal cycles, denoted by $\bm{e}$ (Sec.~\ref{sec:cycle}). Considering reversible reactions is convenient for discussing conformality and thermodynamic bounds. For numerical enumeration, however, it is advantageous to exploit the standard formulation used in metabolic network analysis, where all reactions are treated as irreversible and EFMs are identified with the extreme rays of a non-negative flux cone \cite{schuster1994ELEMENTARY,schilling1999Metabolic,muller2016Elementary}.

For a fully irreversible network with internal stoichiometric matrix $\bS^{X,\mathrm{irr}}$, the steady-state flux cone is
\begin{equation}
  \mathcal{C}^{\mathrm{irr}} \equiv \big\{ \bm{v} \in \mathbb{R}^{N_R}_{\ge 0} \,\big|\, \bS^{X,\mathrm{irr}} \bm{v} = \bm{0} \big\},
\end{equation}
and EFMs are in one-to-one correspondence with the extreme rays of $\mathcal{C}^{\mathrm{irr}}$ \cite{schuster1994ELEMENTARY,schilling1999Metabolic,muller2016Elementary}.
To connect this picture with our reversible setting, we introduce a split representation of the CRN: each reversible reaction is replaced by two irreversible reactions, forward and backward, leading to the split stoichiometric matrix
\begin{equation}
  \bS^{X,\text{split}} \equiv [\bS^X,-\bS^X] \in \mathbb{R}^{N_X \times 2N_R},
\end{equation}
and to the associated non-negative flux cone
\begin{equation}
  \mathcal{C}_{\text{split}} \equiv
  \big\{ \bm{w} \in \mathbb{R}_{\ge 0}^{2N_R} \,\big|\, \bS^{X,\text{split}} \bm{w} = \bm{0} \big\}.
\end{equation}
We denote by $\mathscr{E}^{\text{split}}$ the set of extreme rays of $\mathcal{C}_{\text{split}}$.
Each ray $\bm{w}\in\mathscr{E}^{\text{split}}$ defines an EFM in the original reversible convention via the projection
\begin{equation}
  v_\rho = \sum_{k:\,\pi(k)=\rho} d(k)\,w_k,
\end{equation}
where $\pi(k)$ maps split reactions back to the original reaction index $\rho$ and $d(k) = \pm 1$ encodes forward/backward orientation.
Rays for which both the forward and the backward branch of the same reaction are active correspond to unphysical cycles and are discarded.
The remaining projected vectors $\bm{e}$ are then normalized to co-prime integers and collected in the set $\mathscr{E}$ used throughout the main text. This step avoids redundancy upon global rescaling. In this way, the reversible EFMs of Sec.~\ref{sec:cycle} are obtained from the extreme rays of a non-negative cone, allowing us to apply standard EFM-identification algorithms as developed in metabolic pathway analysis and polyhedral geometry \cite{schuster1994ELEMENTARY,schilling1999Metabolic,muller2016Elementary,bilancioni2025Gearsa}. 

In practice, we employ two complementary exact algorithms to enumerate $\mathscr{E}^{\text{split}}$, and hence $\mathscr{E}$.

\paragraph*{Subset-based circuit enumeration (combinatorial method).}
This method directly searches for conformal circuits in $\bS^{X,\text{split}}$, i.e., minimal subsets of columns whose nullspace is one-dimensional and spanned by a strictly positive vector.
More precisely, let $r = \mathrm{rank}(\bS^{X,\text{split}})$.
We iterate over all subsets of reactions $I \subseteq \{1,\dots,2N_R\}$ with $2 \le |I| \le r+1$, and denote by $\bS^{X,\text{split}}_{I}$ the submatrix obtained by restricting $\bS^{X,\text{split}}$ to the columns in $I$.
We compute the nullspace of $\bS^{X,\text{split}}_{I}$ and retain those subsets for which:
(i) the nullspace is one-dimensional;
(ii) the corresponding null vector has full support on $I$;
and (iii) all entries can be chosen non-negative.
Each conformal circuit obtained in this way yields an extreme ray of $\mathcal{C}_{\text{split}}$ and, after projection to the original coordinates, an element of $\mathscr{E}^{\text{split}}$, and hence an EFM in $\mathscr{E}$.
This approach is conceptually simple and has no external dependencies, but the number of candidate subsets grows combinatorially with the rank of $\bS^{X,\text{split}}$. We therefore use it mainly for small networks or as a cross-check of the geometric method.

\paragraph*{Polyhedral extreme-ray enumeration (geometric method).}
For medium-sized networks, we rely on a geometric algorithm that computes the extreme rays of $\mathcal{C}_{\text{split}}$ using tools from polyhedral geometry.
We represent $\mathcal{C}_{\text{split}}$ as a pointed polyhedral cone defined by the homogeneous constraints $\bS^{X,\text{split}} \bm{w} = \bm{0}$ and $\bm{w} \ge 0$, and use a standard vertex/ray enumeration routine (implemented via the Parma Polyhedra Library \cite{bagnara2006Parma}) to obtain its minimal generating rays.
This approach is output-sensitive: the runtime scales essentially with the number of EFMs rather than with the total number of column subsets to be tested.
In all examples presented in the main text, both algorithms produce identical EFM sets $\mathscr{E}$; however, the geometric extreme-ray method remains dramatically faster for networks with more than $\mathcal{O}(10)$ reactions.
We stress that a variety of other exact and heuristic strategies exist for EFM and extreme-ray computation---including improved polyhedral methods, network decompositions, and constraint-based relaxations---which can further accelerate large-scale applications \cite{schilling1999Metabolic,gerstl2015TEFMA,terzer2008Largescale}.

\begin{figure}[t!]
    \vspace{0.4cm}
  \includegraphics[width=1\columnwidth]{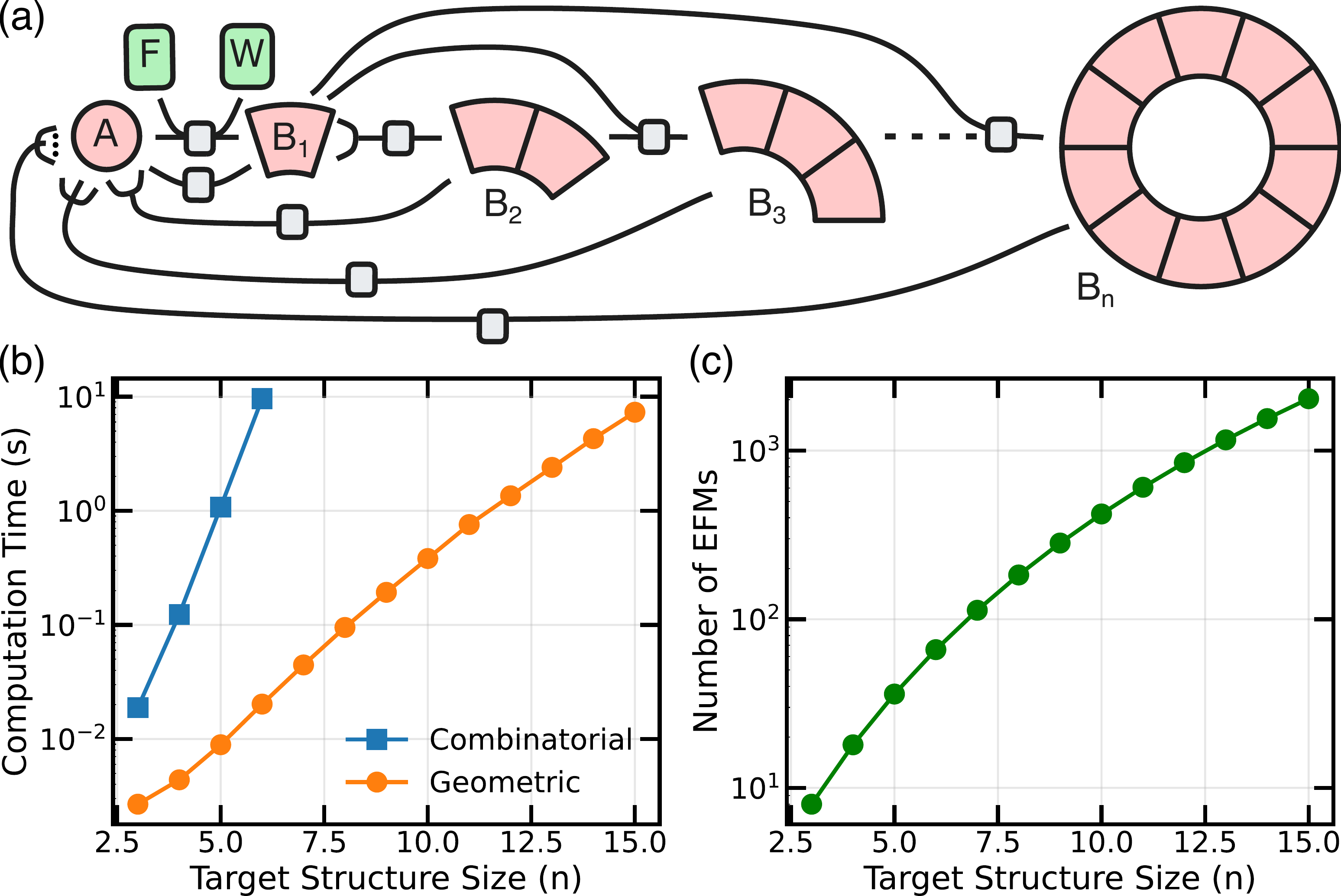}
  \caption{\textbf{Scaling of EFM enumeration for a modular self-assembly network.}
  (a) Reaction scheme of the modular self-assembly network. An inactive monomer $A$ is converted to an activated monomer $B_1$ via a fuel-driven reaction; activated monomers then assemble into aggregates $B_m$ up to size $n$. 
  All intermediate structures undergo spontaneous degradation back to the inactive monomer $A$.
  (b) Computation time as a function of the target structure size $n$ on a logarithmic scale, comparing the combinatorial circuit enumeration (squares) with the geometric extreme-ray method (circles).
  (c) Number of EFMs as a function of $n$ on a logarithmic scale.}
  \label{fig:EFM_search}
\end{figure}

\paragraph*{Application to enlarged assembly space.} To illustrate the computational complexity of EFM enumeration and to demonstrate that our implementation scales to moderately large networks, we consider a self-assembly process analogous to the model discussed in Sec.~\ref{sec:self-assembly}, now generalized to an arbitrary maximal structure size $n$.
The network consists of an inactive monomer $A$, an activated monomer $B_1$, and aggregates $B_m$ of size $m=2,\dots,n$, driven by a chemostatted fuel--waste pair $(F,W)$, as illustrated in Fig.~\ref{fig:EFM_search}. The corresponding reactions are:
\begin{equation}
\begin{aligned}
F + A &\xrightleftharpoons[k_{f-}]{k_{f+}} B_1 + W,\\
B_1 + B_m &\xrightleftharpoons[k_{a-}]{k_{a+}} B_{m+1},\qquad m=1,\dots,n-1,\\
B_m &\xrightleftharpoons[k_{d-}]{k_{d+}} m A,\qquad m=1,\dots,n,
\end{aligned}
\end{equation}
The first reaction is the fuel-driven activation of an inactive monomer $A$ into $B_1$ (powered by $F$ to $W$ conversion); the second describes stepwise assembly, where $B_1$ adds to $B_m$ to form $B_{m+1}$; the third encodes spontaneous disassembly of any aggregate $B_m$ back into $m$ copies of $A$, modeling degradation of intermediate structures. We regard $B_n$ as the target assembled structure, so that increasing $n$ increases both the target size and the size of the underlying reaction network, with the number of reactions in $\bS^X$ equal to $2n$.

For each value of $n$, we construct $\bS^X$, build the split matrix $\bS^{X,\text{split}}$, and compute the full EFM set $\mathscr{E}$ using both the combinatorial and the geometric extreme-ray methods. Figure~\ref{fig:EFM_search}b reports the computation time as a function of the target structure size $n$ on a logarithmic scale, comparing the combinatorial circuit enumeration (squares) with the geometric extreme-ray method (circles). Both curves display the expected {asymptotic} exponential growth with $n$, reflecting the combinatorial explosion of EFMs, but the geometric method remains orders of magnitude faster and reaches values of $n$ for which the combinatorial search becomes impractical.
In Fig.~\ref{fig:EFM_search}c, we show the number of EFMs found as a function of $n$, which also displays approximately exponential growth {for sufficiently large $n$}.
These benchmarks indicate that, while exact EFM enumeration is inherently hard for very large networks, our extreme-ray implementation is sufficient to treat the small and medium-sized CRNs used in this work and to provide quantitative insight into the structure of their thermodynamic space.
In future work, one could combine our thermodynamic-space analysis with more advanced EFM solvers developed in metabolic network theory to push these computations towards larger biochemical systems.

\section{Introducing TACOS}\label{sec:TACOS}

To facilitate the application of the theoretical framework developed in this work, we have created TACOS ({\bf T}hermodynamically {\bf A}ccessible {\bf C}oncentrations of {\bf O}ut-of-equilibrium {\bf S}tationary chemical reaction networks), an open-source Python package that computes the thermodynamic space of small- to medium-size CRNs.

Given the internal stoichiometric matrix $\bS^X$, the standard chemical potentials $\bm{\mu}^0_X$ of internal species, and the external affinity contributions $A^Y_\rho$ encoding the non-equilibrium driving, TACOS enumerates all elementary flux modes using the geometric extreme-ray algorithm described in Appendix~\ref{sec:EFM_enumeration}. From this enumeration, the package derives the affinity bounds of Eq.~\eqref{eq:affinity_bound_1} and the concentration log-ratio bounds of Eq.~\eqref{eq:eq_constant_bound_0} for each reaction in the network. The package also supports chemical probe analysis, enabling the evaluation of thermodynamic constraints on any species interconversion compatible with the conservation laws, even when such interconversions are not explicitly present as reactions in the original network.

TACOS is available under the MIT license at \texttt{https://github.com/Shiling42/TACOS}.

\bibliography{Refs}
\end{document}